\documentclass[mnsc,nonblindrev]{informs3} 

\OneAndAHalfSpacedXI 
\usepackage{hyperref}

\usepackage{endnotes}
\let\footnote=\endnote

 \newcommand{\E}{\mathbb E}
\usepackage{natbib}
 \bibpunct[, ]{(}{)}{,}{a}{}{,}%
 \def\bibfont{\small}%
 \newcommand{\Var}{\mathrm{Var}}

\usepackage{algorithm}
\usepackage{algorithmic}
\usepackage{subfigure}
\usepackage{makecell}

\usepackage{ulem}
\usepackage{booktabs}
\usepackage{multirow}
\usepackage{comment}

\usepackage{hyperref}

\usepackage{xcolor}
\hypersetup{
    colorlinks,
    linkcolor={red!50!black},
    citecolor={blue!50!black},
    urlcolor={blue!80!black}
}
\TheoremsNumberedThrough     
\ECRepeatTheorems

\EquationsNumberedThrough    

\newenvironment{assumption'}[1]
  {\renewcommand{\theassumption}{\arabic{assumption}$'$}%
   \addtocounter{assumption}{-1}%
   \begin{assumption}}
  {\end{assumption}}

\usepackage{amsmath}
\usepackage{amsfonts}
\usepackage{amssymb}
\usepackage{dsfont}
\newenvironment{alg'}[1]
  {\renewcommand{\thealgorithm}{\ref{#1}$'$}%
   \addtocounter{algorithm}{-1}%
   \begin{breakablealgorithm}}
  {\end{breakablealgorithm}}

\newenvironment{lemma'}[1]
  {\renewcommand{\thelemma}{\ref{#1}$'$}%
   \addtocounter{lemma}{-1}%
   \begin{lemma}}
  {\end{lemma}}

\begin{document}



\RUNTITLE{Joint Optimization and Statistical Inference for Zero-th Order Simulation Optimization}

\TITLE{Joint Optimization and Statistical Inference for Zero-th Order Simulation Optimization}

\ARTICLEAUTHORS{%
\AUTHOR{Yuhang Wu, Zeyu Zheng}
\AFF{ Department of Industrial Engineering and Operations Research, University of California, Berkeley, CA}
\AUTHOR{Yingfei Wang}
\AFF{Foster School of Business, University of Washington, Seattle, Washington
}
\AUTHOR{Guangyu Zhang, Zuohua Zhang, Chu Wang}
 \AFF{Amazon.com Inc, Seattle, Washington}
 }

\ABSTRACT{%
We consider stochastic optimization problems with the dual tasks of (i) effectively finding the optimizer and (ii) reliably conducting statistical inference for the optimal objective function value. We find that classical simulation optimization and stochastic optimization algorithms, despite of their fast convergence rates to the optimizer under strong convexity assumptions, may not come with a valid central limit theorem (CLT) with a vanishing bias. This non-vanishing bias can harm statistical inference and the construction of asymptotically valid confidence intervals. We fix this issue by providing a new stochastic optimization algorithm that on one hand maintains the same fast convergence rate and on the other hand permits the establishment of a valid CLT with vanishing bias. We discuss practical implementations of the proposed algorithm and conduct numerical experiments to illustrate the theoretical findings. 
}%


%

\maketitle

%


\section{Introduction}

We consider stochastic optimization problems with the dual tasks of (i) effectively finding the optimizer and (ii) reliably conducting statistical inference for the optimal objective function value. In particular, we consider in this work the so-called \textit{zero-th order stochastic optimization} or \textit{black-box stochastic optimization} with continuous decision variable. The term ``zero-th order" and ``black-box" refer to scenarios where only noisy function value evaluations are available, but no gradient information of the function value is directly available. That is, if one wants to obtain gradient information, it would be necessary to use the finite-difference method and its variants to construct gradient estimators; see Chapter VIII of \cite{asmussen2007stochastic} for reference.

Within the aforementioned dual tasks, the first task of effectively finding the optimizer has been well studied in the simulation optimization and stochastic optimization literature.  To put the discussion in a bit more context, take a simple form $\theta^* = \argmin_{\theta}\mu(\theta)$ for example. In a ``zero-th order" stochastic optimization, for each sample, one can set a choice of value for $\theta$, and observe a noisy evaluation of $\mu(\theta)$. In the literature, algorithms such as the Kiefer–Wolfowitz algorithm have done a provably optimal job in effectively finding the optimizer $\theta^*$, say using a given budget of $T$ samples. If we denote the estimator for $\theta^*$ identified by an algorithm after $T$ samples as $\hat{\theta}_T$, the literature has largely focused on analyzing the convergence rate of $\hat{\theta}_T$ to $\theta^*$ (and the associated $\mu(\hat{\theta}_T)$ to $\mu(\theta^*)$) as $T$ gets large. Part of the literature also develops other large-sample properties for $\hat{\theta}_T$. 

In addition to this aforementioned well-studied first task, we consider the simultaneous execution of the second task --- reliably conducting statistical inference for the optimal objective function value. This second task pursues that after $T$ samples are spent, in addition to obtaining a good $\hat{\theta}_T$, we also want to simultaneously provide a valid confidence interval for $\mu(\theta^*)$. This second task involves providing an estimator for $\mu(\theta^*)$ using the $T$ samples, building a valid central limit theorem (CLT), and showing that the CLT has vanishing bias at the scale of $T^{-1/2}$. Only knowing the convergence rate of $\mu(\hat{\theta}_T)$ to $\mu(\theta^*)$ would not suffice for this second task, because on one hand, $\mu(\cdot)$ (and in particular $\mu(\hat{\theta}_T)$) is not known and the estimation of which would need a sufficient amount of appropriate samples, and on the other hand showing CLT with vanishing bias for the estimator would encounter additional challenges at the relevant $T^{-1/2}$ scale.

We hope to provide some discussions on why this second task of conducting statistical inference for $\mu(\theta^*)$ is of interest in some applications. One of such applications arises from adaptive A/B tests with treatment parameter optimization. For a range of A/B tests implemented in practice, the treatment plan is not a single plan, but is instead a infinite continuum of plans indexed by a continuous parameter. The different value of such parameter can lead to different expected outcome of the treatment plan. When the parameter $\theta$ takes continuous value, the treatment plan itself effectively has a continuum infinite number of variants, labeled by each choice of value for $\theta$. We call such parameter that may affect the treatment plan's expected outcome as \textit{treatment parameter}.
For adaptive A/B tests with treatment parameter optimization, both tasks are of strong interests, including finding the optimal treatment parameter and statistically asserting that whether the treatment effect under the estimated optimal treatment parameter is larger than some threshold value. 

For A/B tests applications, the emergence of a non-vanishing bias in the CLT is a critical issue and can lead to wrong decisions. On one hand, with a non-vanishing bias, the conclusion on whether an optimized treatment is better than the control would be unreliable, because the bias can flip the sign of the difference between treatment and control. On the other hand, it is generally the case that if there is a non-vanishing bias, there would be no effective way of knowing how large such bias is, because knowing the bias would need the knowledge of the unknown true value to be estimated in the first place. This notion is also presented in the Monte Carlo simulation area, where there are usually multiple ways to estimate the variance, but rather limited ways to nail down what exactly the bias is, if there is a bias. 

In addition to the A/B tests settings where the statistical inference task for the unknown optimal objective function value is a natural need, this task may also be relevant to other simulation optimization and stochastic optimization tasks. If the application goal is to solely find the optimizer (i.e., the best decision/action to take), then such statistical inference task would not be necessary to focus on. However, if one also concerns about whether the objective function value under the optimal decision is better than some threshold value, then this statistical inference task would be of direct relevance.

We provide a brief discussion on the connection and trade-off between the two tasks --- abbreviated and referred to as optimization and statistical inference.  On the one hand, we need to find a good estimator of $\theta^*$, and this goal can generally be effectively achieved by standard zeroth-order simulation/stochastic optimization algorithms  under convexity assumptions. On the other hand, samples adaptively acquired by standard algorithms generally focus on construct gradient estimators, whereas these samples may not be sufficiently representative to provide a good estimator for the optimal objective value $\mu(\theta^*)$, which is crucially needed for statistical inference. We provide accurate analysis and discussion in the main body of the work.

The results of this work can be summarized as follows.

\begin{enumerate}
    \item We consider and formulate the dual tasks of optimization and statistical inference, which consists of (i) solving a simulation optimization problem with only noisy function value evaluations (i.e., zero-th order) and (ii) a need to establish central limit theorems for the estimated optimal function value. We argue that even though classical zeroth-order simulation optimization or stochastic optimization algorithms can be used to asymptotically find the optimizer at fast rates under a strongly convex assumption on the objective function, they do not come up with a central limit theorem that can be used for valid statistical inference due to a non-vanishing bias. Such bias is generally difficult to quantify, if not impossible at all.   
    \item 
    We fix this gap by providing a new optimization algorithm that on the one hand maintains the same fast convergence rate and on the other hand permits the establishment of a valid central limit theorem with vanishing bias that can be used to construct asymptotically valid confidence intervals. The new algorithm strategically draws four noisy function values to construct estimators for the gradient and for the objective function value, instead of the classical finite difference gradient estimator based on two draws of noisy function values. 
    
    \item We prove under a strong convexity assumption that our proposed algorithm enjoys a convergence rate of $O(T^{-\frac{1}{3}+\varepsilon})$ for any arbitrary small $\varepsilon>0$ on the estimator of the optimizer, and simultaneously enjoys an almost optimal central limit theorem (CLT) for the estimated optimal objective function value, a CLT that is almost as good as if the optimal parameter were known in advance. We conduct numerical experiments to illustrate the theoretical findings. 
\end{enumerate}

\subsection{Connections to Literature}

Our work is closely related to work on simulation optimization with continuous variable. There is a rich literature on this topic, including   \cite{l1994stochastic,hu2008model,fu2008some,andradottir2010adaptive,pasupathy2013simulation, hashemi2014adaptive,zhou2014simulation,zhou2017gradient,fan2018surrogate,kiatsupaibul2018single,sun2018gaussian,hunter2019introduction,hong2020finite,zhang2021actor,hong2021surrogate}. The setting considered by our work naturally forms a simulation optimization problem where only noisy function value evaluation is available. In our setting, each sample comes either from feeding one of the plans to a customer or from a black-box simulator where the internal structure is not easy to change. Therefore, in our setting, it is not possible to use, for example, infinitesimal perturbation analysis to conveniently obtain a gradient estimator. This is because the outcome observed from each sample can be viewed as an outcome observed from a black-box noisy function oracle. There is not much we can do to look into the black box and obtain more information. Another feature of our setting is that it is difficult, if not impossible, to apply common random numbers.

Our work is also connected to the stochastic approximation literature; see \cite{robbins1951stochastic, kiefer1952stochastic,spall1992multivariate,l1998budget,kushner2003stochastic,kim2015guide,he2021adaptive,hu2022stochastic} among others. Stochastic approximation problems generally aim to adaptively find the root of an equation in some region of the variable. Stochastic approximation and simulation/stochastic optimization are tightly connected. One way to connect is that some simulation/optimization problems can be formulated into finding the root of the equation where the gradient is set to zero. Our work is especially connected to \cite{l1998budget}, who consider constructing central limit theorems for stochastic approximation problems with a given budget. Even though the setting is different, our central limit theorem proof is partially inspired by \cite{l1998budget}.

Our work is connected to the literature of stochastic zeroth-order optimization; see \cite{ghadimi2013stochastic,duchi2015optimal,nesterov2017random,liu2018zeroth,balasubramanian2018zeroth,ajalloeian2020convergence} among others. The commonality in our work is that we also need to use zeroth-order noisy function evaluations to construct gradient estimator and do gradient search for optimizing the parameter. Our work additionally provides statistical inference results in addition to the optimization results. Our work is also connected to \cite{xie2016bayesian, wu2017bayesian, letham2019constrained,wang2020parallel,balandat2020botorch}; they consider the problem of adaptively optimizing a continuous parameter of the treatment plan, under a Bayesian framework. Our work is different in three folds. Our algorithm takes a frequentist view, constructs gradient information using noisy function value observations, and does gradient search, which is different from the Bayesian approach. Our work develops a central limit theorem (CLT) that can be directly used for large-sample statistical inference, while the above mentioned work does not emphasize on CLT. 

In addition, our work is connected to and different from the literature of post-selection inference, see for example \cite{luedtke2016statistical, nie2018adaptively, hadad2021confidence,andrews2024inference}. For the literature of post-selection inference, the majority of work focuses on resolving the so-called ``winner's curse" or using data from adaptive algorithms (e.g., bandit algorithm) to perform inference. The connection of our work to this literature is that we also have the step of using post-algorithm data to perform statistical inference. The difference of our work compared to this literature is mainly two-fold: first, our work considers a stochastic optimization problem with continuous variable, different from most of the selection problem with a finite number of choices; second, our work is not passively conducting inference after an adaptive optimization algorithm is run, but also actively changing the optimization algorithm to fulfill the dual tasks. 

\textbf{Organization.}  In Section \ref{sec:alg}, we describe the problem setting for \textit{joint parameter optimization and statistical inference} and provide a new algorithm design. In Section \ref{sec:theory}, we provide theoretical guarantees and central limit theorem for the proposed algorithm. Section \ref{sec:numerical_exp} presents numerical experiments. Section \ref{sec:conclusion} concludes.

\section{Problem setting and Algorithm}\label{sec:alg}
\subsection{Problem setting}

We write the problem setting for \textit{joint parameter optimization and statistical inference}. Suppose the outcome of the model we are interested in is given by 
\begin{align}
    Y(\theta) = \mu(\theta) + \sigma(\theta)\cdot \epsilon_{\theta}, \quad \theta  \in \Theta.
\end{align}
Here for simplicity we assume $\Theta=[0,1]^d$. $\mu(\theta)$ is the expected outcome given the parameter $\theta$, the term $\sigma^2(\theta)$ is the variance of the outcome, and $\epsilon_{\theta}$ is a random variable with mean zero and variance one. For different choices of value for $\theta$, the probability distribution of $\epsilon_{\theta}$ is allowed to be different.

 \textbf{Our feasible actions: adaptive parameter assignment.} 
 For each $i=1,\cdots, T$, we need to determine a choice of $\theta_{i}$ based on the values of $\{y_j,\theta_j\}_{j=1}^{i-1}$. Here $y_j$ is a realization of $Y(\theta_j)$ we sampled. After $\theta_i$ is chosen, we observe a realization of $Y(\theta_i)$ given by $y_i$. We assume all $Y(\theta_i)$ are independent.

\textbf{Our tasks.} 
There are two simultaneous goals. (i) For a given sample size $T$, we hope to recommend a good optimizer $\hat{\theta}_T$ to minimize $\mu(\theta)$; (ii) We want to provide valid inference for $\mu({\theta}^*)$. Specifically, we want to prove a central limit theorem (CLT) for the estimator, with vanishing bias at the scale of $T^{-1/2}$. The asymptotic distribution in the CLT is then used to construct valid confidence interval and hypothesis test procedures if needed.

\subsection{Our algorithm}

In this section, we propose an algorithm that one can use to the two tasks described in the previous section. At the first glance, a straightforward idea to deal with this implicit trade-off is a ``search-then-evaluate" strategy, that is, we first use part (say half) of samples to search for a good estimator $\hat\theta$ for the optimizer $\theta^*$, then we fix this $\hat{\theta}$ and use the remaining samples to draw noisy observations of $\mu(\hat\theta)$. This idea works in the sense that it indeed recommends a $\hat{\theta}$ that converges to optimizer and also provides a CLT centered at $\mu(\hat{\theta})$ when $T\to+\infty$. However, it can suffer from inefficiency and slower rates in both the searching part and evaluation part. As for the searching procedure, since we only use part of all samples, the estimator for the optimizer may not be good enough compared with those algorithms searching with all samples. For the evaluation part, using part of samples can give a larger variance in the CLT compared with using all samples, which then leads to low statistical power for testing and inference. We provide some numerical results to illustrate this point in the Appendix. Thus, we accordingly seek for better remedies.

In this work, we propose an algorithm that adaptively decides how to use the $T$ samples to search over different  choices of value for the parameter $\theta$, with the goal to fulfill the dual optimization task and statistical inference task. Our algorithm, as will be proved later, enjoy the following properties: 

\begin{enumerate}
    \item[(i).] It uses \textit{all} $T$ samples to do stochastic optimization and has a fast convergence rate as $\E\Vert \hat{\theta}_T-\theta^*\Vert_2=O(T^{-\frac{1}{3}+\eta})$ for arbitrarily small $\eta>0$.
    \item[(ii).] For our recommendation $\hat{\theta}$, it does not waste any samples, in the sense that it uses \textit{all} $T$ samples to construct an estimator of $\mu(\hat\theta)$ and no sample is only used in the optimization task but not the inference task. 
    \item[(iii.)] A CLT holds for our estimator of $\mu(\hat\theta)$ with nearly optimal asymptotic variance compared with the CLT where hypothetically ${\theta}^*$ was known and we sample $Y({\theta}^*)$ for $T$ times to perform inference for $\mu({\theta}^*)$. 
\end{enumerate}

The flow of our algorithm is as follows. Suppose we have in total $T$ samples, and we begin with an initial point $\theta^0$. In $t$-th iteration, to find a descending direction, we randomly sample a vector $w^t$ uniformly from a unit sphere surface $B_d(0,1)=\{x\in[-1,1]^d \big\vert \Vert x\Vert_2=1\}$, and define following four points:
\begin{equation}\label{theta_plus}
\theta^+=\theta^t+c_tw^t,\,\,\theta^-=\theta^t-c_tw^t,
\end{equation}
\begin{equation}\label{theta_plusplus}
\theta^{++}=\theta^t+kc_tw^t,\,\,\theta^{--}=\theta^t-kc_tw^t.
\end{equation}
We omit the projection operator back into the feasible region $\Theta$ if some of the four points exceed the boundary of $\Theta$. Here, $k>1$ is an algorithm parameter, and we will discuss how to choose this parameter in a principled way later. The parameters $\{c_t\}$ are half lengths of finite difference which will also be determined later.

Note that there is no first-order oracle to draw gradient estimator directly. In order to estimate the gradient through some numerical scheme, we will first estimate $\mu(\theta^+),\mu(\theta^-),\mu(\theta^{++}),\mu(\theta^{--})$. For some predetermined $m_1,m_2\in \mathbb{N}_+$, let $m=m_1+m_2$, without loss of generality we assume that $\frac{T}{2m}\in\mathbb{N}$. Now, we sample the random variable $Y(\theta^+)$ for $m_1$ times to obtain its realizations $y_{2m(t-1)+i} (i=1,\cdots,m_1)$, and also sample $Y(\theta^-)$ for $m_1$ times to obtain $y_{2m(t-1)+i}(i=m_1+1,\cdots,2m_1)$. Similarly, we can also generate $y_{2m(t-1)+2m_1+j}(j=1,\cdots,m_2)$ as realizations of $Y(\theta^{++})$ and $y_{2m(t-1)+2m_1+j}(j=m_2+1,\cdots,2m_2)$ as realizations of $Y(\theta^{--})$. We then estimate $\mu(\theta^+)$, $\mu(\theta^-)$, $\mu(\theta^{++})$, $\mu(\theta^{--})$ through sample mean as follows:
\begin{equation}\label{eq:hat_mu_plus}
\begin{aligned}
    \bar{\mu}(\theta^+)&=\sum_{i=1 }^{m_1}\frac{y_{2m(t-1)+i}}{m_1}, \quad
    \bar{\mu}(\theta^-)&=\sum_{i=m_1+1}^{2m_1}\frac{y_{2m(t-1)+i}}{m_1};
\end{aligned}
\end{equation}

\begin{equation}\label{eq:hat_mu_plusplus}
\begin{aligned}
    \bar{\mu}(\theta^{++})&=\sum_{j=1 }^{m_2}\frac{y_{2m(t-1)+2m_1+j}}{m_2}, \quad
    \bar{\mu}(\theta^{--})&=\sum_{j=m_2+1}^{2m_2}\frac{y_{2m(t-1)+2m_1+j}}{m_2}.
\end{aligned}
\end{equation}

Next, the directional derivative at $\theta^t$ w.r.t. $w_t$ is estimated by:
\begin{equation}\label{eq:grad}
    g^t = \frac{-\bar{\mu}(\theta^{++})+k^3\bar{\mu}(\theta^{+})-k^3\bar{\mu}(\theta^{-})+\bar{\mu}(\theta^{--})}{2k(k^2-1)c_t}w^t,
\end{equation}
and an estimator of $\mu(\theta^t)$ is constructed by:
\begin{equation}\label{eq:value}
    \hat{\mu}_t=\frac{-\bar{\mu}(\theta^{++})+k^2\bar{\mu}(\theta^{+})+k^2\bar{\mu}(\theta^{-})-\bar{\mu}(\theta^{--})}{2(k^2-1)}.
\end{equation}
Finally, we update $$\theta^{t+1}\leftarrow\theta^t-\alpha_tg^t$$
for some step size $\alpha_t$ to finish the $t$-th iteration. It ends up recommending a choice of parameter given by $\hat{\theta}_T$; the algorithm also provides an estimator $\hat{\mu}(\hat{\theta}_T)$ for the (unknown) expected outcome $\mu(\hat{\theta}_T)$ on the recommended choice of parameter. The complete algorithm is provided in Algorithm \ref{alg:xxx}.

Here the gradient estimator \eqref{eq:grad} and function value estimator \eqref{eq:value} may seem not intuitive at first glance. They use four points instead of two points in standard finite difference methods. We provide some analysis as follows, and discuss why the classical two-point method cannot work in Section \ref{subsec:grad}. For \eqref{eq:value}, if we consider $\mu(\cdot)$ to be third-order differentiable, we have for small $h>0$ that
$$\mu(\theta+h)+\mu(\theta-h)=2\mu(\theta)+h^T\nabla^2\mu(\theta)h+O(\Vert h\Vert_2^3),$$
then plug in and we obtain:
$$\mu(\theta^+)+\mu(\theta^-)=2\mu(\theta^t)+c_t^2(w^t)^\top\nabla^2\mu(\theta)2w^t+O(c_t^3),$$
and
$$\mu(\theta^{++})+\mu(\theta^{--})=2\mu(\theta^t)+k^2c_t^2(w^t)^\top\nabla^2\mu(\theta^t)w^t+O(c_t^3).$$
From above two equations we have: $$
{\mu}(\theta^t)=\frac{-{\mu}(\theta^{++})+k^2{\mu}(\theta^{+})+k^2\mu(\theta^-)-\mu(\theta^{--})}{2(k^2-1)}+O(c_t^3),$$
which implies that \eqref{eq:value} gives an estimator of the function value with third order error. Thanks to this error order, the ``not intuitive" construction for this estimator can achieve a high-enough accuracy to later facilitate the establishment of a CLT with vanishing bias. 

As for the gradient estimator construction in \eqref{eq:grad}, it uses four points instead of two points to construct the finite difference gradient estimator. We can show that the four-point construction \eqref{eq:grad} gives a better estimator compared to the standard two-point construction when stronger smoothness condition holds. In fact, if $\mu(\cdot)$ is fourth-order differentiable, then by fourth-order Taylor expansion we can see that
$$
\begin{aligned}
    \nabla\mu(\theta^t)^\top w^t =& \frac{-{\mu}(\theta^{++})+k^3{\mu}(\theta^{+})-k^3{\mu}(\theta^{-})+{\mu}(\theta^{--})}{2k(k^2-1)c_t}+O(c_t^4),
\end{aligned}
$$
which gives an estimator with fourth-order error, while the standard two-point method has second-order error. If one does not believe that $\mu(\cdot)$ is smooth enough, then the above gradient estimator is not necessary and can have a larger variance, so one can turn to a standard gradient estimator given by:
    \begin{equation}
        g^t_0=\left(\frac{\bar{\mu}(\theta^{++})-\bar{\mu}(\theta^{--})}{2kc_t}+\frac{\bar{\mu}(\theta^{+})-\bar{\mu}(\theta^{-})}{2c_t}\right)w^t,
    \end{equation}
which is just an estimator based on the finite difference method.

\begin{algorithm}[!t]
\caption{Zeroth-order optimization for the dual task}\label{alg:xxx} 
\begin{algorithmic}[1]
\STATE \textbf{Input:} initial point $\theta^0 \in [0,1]^d$,  parameters $k>1$, $\beta\in(0,1)$, step sizes $\{\alpha_t\}$, subsample sizes $m_1,m_2$, half lengths of finite difference interval $\{c_t\}$, and total samples $T$ that satisfies $\frac{T}{2(m_1+m_2)}\in\mathbb{N}$.

\FOR{$t = 1, 2, \ldots, \frac{T}{2m}$}
\STATE Sample $w^t$ uniformly from a unit sphere $B_d(0,1)$.
\STATE Calculate $\theta^+,\theta^-,\theta^{++},\theta^{--}$ through \eqref{theta_plus} and \eqref{theta_plusplus}.
\FOR{$i=1,\ldots,2m_1$}
\STATE Sample $y_{2(m_1+m_2)t+i}$ as a realization of $Y(\theta^+)$ if $1\leq i\leq m_1$ otherwise $Y(\theta^-)$;
\ENDFOR
\FOR{$j=1,\ldots,2m_2$}
\STATE Sample $y_{2(m_1+m_2)t+2m_1+j}$ as a realization of $Y(\theta^{++})$ if $1\leq j\leq m_2$ otherwise $Y(\theta^{--})$;
\ENDFOR
\STATE Compute $\bar{\mu}(\theta^+),\bar{\mu}(\theta^+),\bar{\mu}(\theta^{++}),\bar{\mu}(\theta^{--})$ through \eqref{eq:hat_mu_plus},\eqref{eq:hat_mu_plusplus}, respectively.
\STATE Compute $g^t$ through \eqref{eq:grad}.
\STATE Compute $\hat{\mu}_t$ through \eqref{eq:value}.
\STATE Update
$\theta^{t+1} \leftarrow \theta^{t} - \alpha_t g^t$.
\ENDFOR
\STATE \textbf{Output:} Take $\hat{\theta}_T=\theta^{T/2m}$ and $\hat{\mu}(\hat{\theta}_T)=\frac{2m}{T}\sum_{t=1}^{T/2m} \hat{\mu}_t. $
\end{algorithmic}
\end{algorithm}

\subsection{The failure of classical finite difference method}\label{subsec:grad}

Before establishing a CLT for $\hat\mu(\hat{\theta}_T)$, we  discuss why the classical two-point finite difference method cannot work well for the dual task of optimization and inference, despite its success when solely addressing the optimization task. 

Suppose we now use the classical two-points approach to optimize $\mu(\theta)$ with only noisy samples available, and further assume $\mu(\theta)$ is convex and second-order continuously differentiable. We consider the use of central FD gradient estimator, which is prevalent in zeroth-order simulation/stochastic optimization literature. Since it is generally hard to give a lower bound on the convergence of an optimization algorithm, we provide following heuristic arguments and also some numerical examples in Section \ref{sec:numerical_exp} to show that it cannot provide valid statistical inference. For the central FD estimator and any sample $\theta^t$
$$
    g^t_\text{FDC}=\frac{\bar\mu(\theta^+)-\bar\mu(\theta^-)}{2c_t}.
$$
Suppose we now replace \eqref{eq:grad} by $g^t_\text{FDC}$, and take $c_t=O(t^{-\nu})$ for some $\nu>0$, then from the literature (e.g. \cite{l1998budget}) we know that the variance of $g^t_\text{FDC}$ is $O(t^{2\nu})$ and $\E\Vert\theta^t-\theta^*\Vert_2^2=O(t^{\min(-1+2\nu,4\nu)})=O(t^{-1+2\nu})$. Since $|\mu(\theta^t)-\mu(\theta^*)|=O(\Vert\theta^t-\theta^*\Vert_2^2)$, then 
$$\sqrt{T}\Big|\E\frac{2m}{T}\sum_{t=1}^{T/(2m)}\mu(\theta^t)-\mu(\theta^*)\Big|=O(T^{-\frac{1}{2}+2\nu}).$$
In addition, now our estimator for $\mu(\theta^t)$ is $\hat{\mu}_{t,\text{FDC}}=\frac{\bar{\mu}(\theta^+)+\bar{\mu}(\theta^-)}{2}$, and under mild conditions, we can show by Taylor expansion that the bias caused by the FD estimator is $\hat{\mu}_{t,\text{FDC}}-\mu(\theta^t)=O(c_t^2)=O(t^{-2\nu})$. Then  $$\sqrt{T}\Big|\E\frac{2m}{T}\sum_{t=1}^{T/(2m)}\hat{\mu}_{t,\text{FDC}}-\E\frac{2m}{T}\sum_{t=1}^{T/(2m)}\mu(\theta^t)\Big|=O(T^{\frac{1}{2}-2\nu}).$$

For valid statistical inference, a CLT with following form with vanishing bias is needed:
$$\sqrt{T}\Big(\frac{2m}{T}\sum_{t=1}^{T/(2m)}\hat\mu_{t,\text{FDC}}-\mu(\theta^*)\Big)\stackrel{d}\to\mathcal{N}(0,\sigma^2).$$
We therefore will expect $\frac{1}{2}-2\nu<0$ and $-\frac{1}{2}+2\nu<0$, which is evidently impossible for the classical two-point gradient estimator based approach. The basic idea of above arguments is that, when we are using the central FD estimator, a choice of small $c_t$ will give large variance and thus lead to slow convergence rate, but a choice of large $c_t$ can lead to a bad estimator of $\hat\mu(\theta^t)$. This trade-off makes standard two-point FD estimator very difficult to assist the dual tasks. In fact, we provide numerical examples in Section \ref{sec:numerical_exp} to show that, even if $\mu(\cdot)$ is simply a quadratic function on $\mathbb{R}$, a valid CLT with vanishing bias will not hold for the central FD estimator.

\subsection{On the choice of parameters}\label{subsec:choice_of_para}
We now discuss how to choose algorithm parameters $m_1,m_2$ and $k$. First, the choice of $m$ will not affect our algorithm too much since we can hedge its effect by tuning the step size, that is, when $m$ is large, then we can increase the step size to balance the effect.

To get an intuition on how to choose $m_1,m_2$ for fixed $k$ when $m$ is given, we begin by considering a simple case that $Y(\theta)\sim \mathcal{N}(0,1)$ for all $\theta$. In this case, by some calculations we have  
$
    \Var(\hat{\mu}_t)=\frac{1}{2(k^2-1)^2}\big(\frac{k^4}{m_1}+\frac{1}{m_2}\big).
$
To get a good estimator $\hat\mu(\hat\theta_T)$, recall that $\hat\mu(\hat\theta_T)=\frac{2m}{T}\sum_{t=1}^{T/(2m)}\hat{\mu}_t$, we would expect that the variance of $\hat\mu_t$ can be minimized. Then by Cauchy inequality we can see that the variance is minimized at $m_1=\frac{k^2}{k^2+1}m$ and $m_2=\frac{1}{k^2+1}m$, and the minimum value is $\frac{k^2+1}{2(k^2-1)m}$. This argument can be easily adapted to the general case and under some assumptions (will be detailed in the next section) we have:
\begin{proposition}\label{prop:var_m}
Suppose $m,k$ are given, under Assumption \ref{assumption:mu(theta)},\ref{assumption:sigma},\ref{assumption:ode},\ref{assumption:noise}, if $\alpha_t=\frac{C_0}{t}$ for some $C_0>0$ and $c_t=\frac{C_1}{t^{\nu}}$ for some $C_1>0$ and $\frac{1}{6}<\nu<\frac{1}{4}$, we have:
\begin{equation}\label{eq:var_m}
    T\cdot \Var(\hat\mu(\hat{\theta}_T))\to \frac{m}{(k^2-1)^2}\big(\frac{k^4}{m_1}+\frac{1}{m_2}\big)\sigma(\theta^*)^2
\end{equation}
with probability $1$ as $T$ tends to infinity.
\end{proposition}
Proposition \ref{prop:var_m} suggests that $m_1$ and $m_2$ shall be chosen to minimize the right hand side of \eqref{eq:var_m}. Then from our discussions above, we can take $m_1=\frac{k^2}{k^2+1}m$ and $m_2=\frac{1}{k^2+1}m$ in general case.

We now discuss the choice of $k$. Proposition \ref{prop:var_m} indicates that the minimized variance is $\frac{(k^2+1)^2\sigma(\theta^*)^2}{T(k^2-1)^2}$, with the ratio $\frac{(k^2+1)^2}{(k^2-1)^2}$ monotonically decreasing as $k$ increases. Consequently, a larger $k$ results in a smaller asymptotic variance. However, employing a large 
$k$ can destabilize the initialization phase of our algorithm. Recall that we need to sample at $\theta^t+kc_tw^t$ and $\theta^t-kc_tw^t$ to estimate the gradient and function value at $\theta^t$, then with a large $k$ these two points can be far from $\theta^t$, which leads to a large estimation error.
 Given this trade-off, we recommend setting $k\in[3,5]$, At $k=5$, the ratio $\frac{(k^2+1)^2}{(k^2-1)^2}\approx 1.17$, suggesting small benefit from further increases in 
$k$ due to the ratio nearing $1$. Thus, the incremental improvement in variance reduction becomes negligible beyond this point.

\section{Theory}\label{sec:theory}
In this section, we state assumptions on the model and provide a central limit theorem for $\hat{\mu}(\hat{\theta}_T)$ give by our algorithm. Here $T,k,m_1,m_2,q$ are all treated as deterministic now. We begin by following four assumptions.
\begin{assumption}\label{assumption:mu(theta)}
For $\theta\in\Theta$, $\mu(\theta)$ is third-order continuously differentiable and $\nabla^2\mu(\theta)$ is positive-definite.
\end{assumption}
\begin{assumption}\label{assumption:sigma}
For $\theta\in\Theta$, $\sigma(\theta)$ is continuous.
\end{assumption}
\begin{assumption}\label{assumption:ode}
The ordinary differential equation (ODE)
$ \dot\theta=\nabla\mu(\theta)$
has a unique solution for each initial condition.
\end{assumption}
\begin{assumption}\label{assumption:noise}
There exists some $a>0$ and $M_1>0$ such that 
$\E|\epsilon_\theta|^{2+a}<M_1,$
for all $\theta\in[0,1]^d$.
\end{assumption}
{\color{black} Assumption \ref{assumption:mu(theta)}, \ref{assumption:sigma} and \ref{assumption:ode} are relatively standard in the literature of convex simulation optimization and stochastic approximation. For example, they are analogies to Assumption (A3)$-$(A5) in \cite{l1998budget}. Assumption \ref{assumption:mu(theta)} requires that $\mu(\cdot)$ is smooth enough for the tractability of zeroth-order gradient estimator and is strongly convex around the optimizer, where the strongly convex assumption is used in many zeroth-order optimization literature that renders fast convergence rate, see for example \cite{l1998budget,shamir2013complexity,nesterov2017random} and \cite{dvurechensky2021accelerated}.} Relaxing strongly convexity to ordinary convexity would lead to a different slower convergence rate and require a different procedure about deriving central limit theorems. We are not sure about how to make this relaxation and hope to leave this to future work. 

Assumption \ref{assumption:sigma} requires that $\sigma(\cdot)$ is continuous. Assumption \ref{assumption:ode} requires the ODE to be ``well behaved" since this ODE is closely related to the behaviour of $\{\theta^t\}$.  Finally, Assumption \ref{assumption:noise} requires a $(2+a)$th-order moment exists uniformly for all $\epsilon_\theta$, which is a usually made assumption in the derivation of central limit theorem.

We first need to prove following theorem for the convergence rate of Algorithm~\ref{alg:xxx}. The proof is displayed in the Appendix.

\begin{theorem}\label{thm:general_error}
Under Assumption \ref{assumption:mu(theta)}, \ref{assumption:sigma} and \ref{assumption:ode}, if $\alpha_t={c_0}t^{-\rho}$ for some $c_0>0$ and $\frac{3}{4}< \rho\leq 1$, $\lim_{t\to\infty}t^{\nu}c_t=c_1$ for some $c_1>0$ and $\frac{\rho}{6}<\nu<\frac{2\rho-1}{4}$, then we have
$\theta^t\to\theta^*$
almost surely as $t\to\infty$, and 
$$\E\Vert{\theta}^t - \theta^*\Vert_2^2=O(t^{-\rho+2\nu})=o(t^{-\frac{1}{2 }}).$$
\end{theorem}
\begin{remark}
Theorem \ref{thm:general_error} gives the convergence rate for the optimizer recommended by our algorithm. By carefully choosing the step sizes $
\{\alpha_t\}$ and parameter $\nu$, the convergence rate is fast enough in the sense that $\E\Vert{\theta}^t - \theta^*\Vert_2^2=o(t^{-\frac{1}{2}})$, so that it does not contribute additional asymptotic variances to the CLT for estimated optimal function value. In addition, the choice of parameters are also crucial in the next theorem. 
\end{remark}

Based on Theorem \ref{thm:general_error}, our central limit theorem for the optimal objective function value estimator $\hat{\mu}(\hat{\theta}_T)$ can be stated as follows:
\begin{theorem}\label{thm:clt_simple}
Assume all the conditions in Theorem \ref{thm:general_error} hold, and in addition Assumption \ref{assumption:noise} is satisfied, then we have:
\[
\sqrt{T}(\hat{\mu}(\hat{\theta}_T) - \mu(\hat\theta_T)) \overset{d.}{\rightarrow} \frac{(k^2+1)\sigma(\theta^*)}{k^2-1} \mathcal{N}(0,1) 
\]
and
\[
\sqrt{T}(\hat{\mu}(\hat{\theta}_T) - \mu(\theta^*)) \overset{d.}{\rightarrow} \frac{(k^2+1)\sigma(\theta^*)}{k^2-1} \mathcal{N}(0,1) 
\]
as $T\to+\infty$, where $$
\hat{\mu}(\hat{\theta}_T) = \frac{2m}{T}\sum_{i=1}^{T/(2m)} \hat{\mu}_i.$$
\end{theorem}
\begin{remark}
Theorem \ref{thm:clt_simple} provides a  CLT with desired order and vanishing bias.  Here, since we can show that $\E\big|\mu(\hat\theta_T)-\mu(\theta^*)\big|=O(T^{-\gamma})$ with $\gamma>\frac{1}{2}$, the confidence intervals for $\mu(\hat{\theta}_T)$ and $\mu(\theta^*)$ are asymptotically the same. Note that the asymptotic variance is $\frac{(k^2+1)^2\sigma(\theta^*)^2}{(k^2-1)^2}$ and as we discussed in Section \ref{subsec:choice_of_para}, this is close to $\sigma^2(\theta^*)$ when $k$ is large, while the latter one is the optimal variance, which means the estimator provided by our algorithm behaves almost as well as the case that $\theta^*$ is known in oracle.
\end{remark}

 Regarding the variance term that is needed in practical use of CLT, we can construct an estimator $\hat{\sigma}(\theta^*)$ for the unknown $\sigma(\theta^*)$ through the sample standard deviation of $y_1,y_{2},\cdots,y_T$, then $\hat{\sigma}(\theta^*)$ is consistent under some additional conditions (details provided in the Appendix), and an asymptotic $95\%$ confidence interval can be constructed as 
\begin{equation}\label{CI}
    \Big[\hat{\mu}(\hat\theta_T)-\frac{1.96(k^2+1)\hat\sigma(\hat{\theta}_T)}{\sqrt{T}(k^2-1)},\hat{\mu}(\hat\theta_T)+\frac{1.96(k^2+1)\hat\sigma(\hat{\theta}_T)}{\sqrt{T}(k^2-1)}\Big].
\end{equation}

\section{Numerical Experiments}\label{sec:numerical_exp}
In this section, we give some synthetic examples to illustrate the performance of our algorithm and also show that the central finite difference (FD) method can not give the desired CLT. Because our goal is illustrate the central limit theorem performance, limiting distribution and asymptotic bias, we need to know the true underlying value to quantify the bias etc. That is why synthetic examples are needed. Real data generally only give us realized sample paths but not underlying true value to illustrate CLT. Yet we can use data to calibrate a synthetic model. 

We consider a synthetic example where the parameter (a relevance parameter in advertisement recommendation) $\theta\in [0,1]$ is one-dimensional, and $\mu(\theta)$ is calibrated from a set of A/B test data in a e-commerce company as follows: $\mu(\theta)=0.02125\theta^2-0.01825\theta-0.0105$. Here $-\mu(\theta)$ denotes the click-through rate (CTR) and $-Y(\theta)$ is a Bernoulli random variable. Then $\sigma(\theta)=\sqrt{-\mu(\theta)(1+\mu(\theta))}$. We hope to note that our work is not intended to propose an algorithm that outperforms all existing algorithms. Instead, we are aiming at one algorithm that on one hand enjoys desirable theoretical convergence rate and on the other hand well facilities statistical inference through establishing a CLT. 

We set $k=3$, $m=50$, $\nu=\frac{1}{5}$, $\theta^0=\frac{1}{2}$, and $c_t=\frac{1}{t^{\nu}}$, $\alpha_t=\frac{30}{t}$, respectively. We repeat our algorithm $1000$ times and report the normalized and centralized estimator. Here, the normalized and centralized estimator is defined as: $$\frac{\sqrt{T}(k^2-1)}{(k^2+1)\sigma(\theta^*)}\big(\hat{\mu}(\hat{\theta}_T)-\mu(\hat{\theta}_T)\big),$$
so it should follow a standard normal distribution according to Theorem \ref{thm:clt_simple}, which has been justified by our results in Section \ref{sec:numerical_exp}.

\begin{figure}[h]
 \centering 
 \subfigure[Estimator distribution based on  Algorithm \ref{alg:xxx}, $T=10^5$]{
 \label{fig:1e5_algo}
 \includegraphics[scale=0.44]{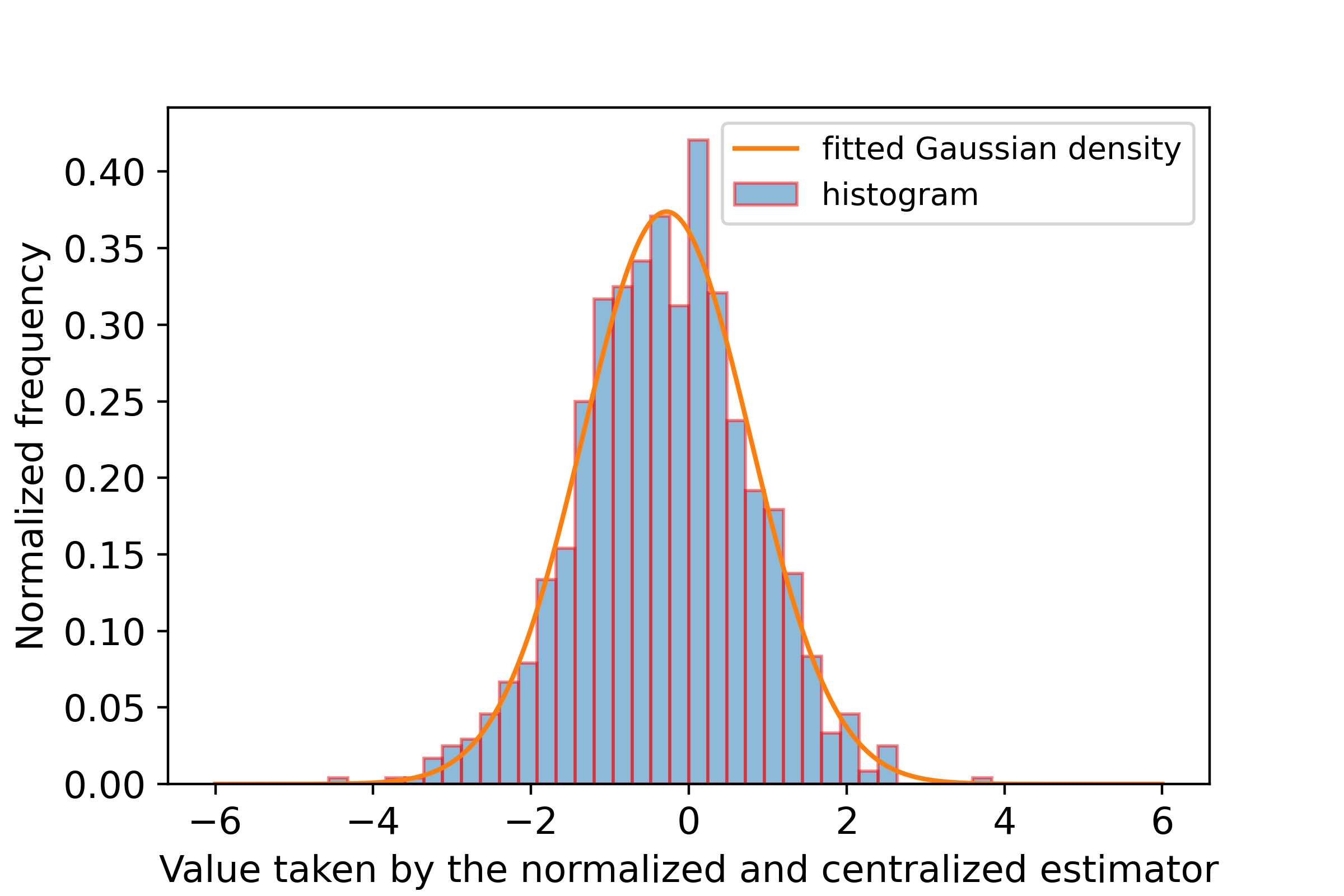}}
 \subfigure[Estimator distribution based on two-point central FD method, $T=10^5$]{
  \label{fig:1e5_naive}
 \includegraphics[scale=0.44]{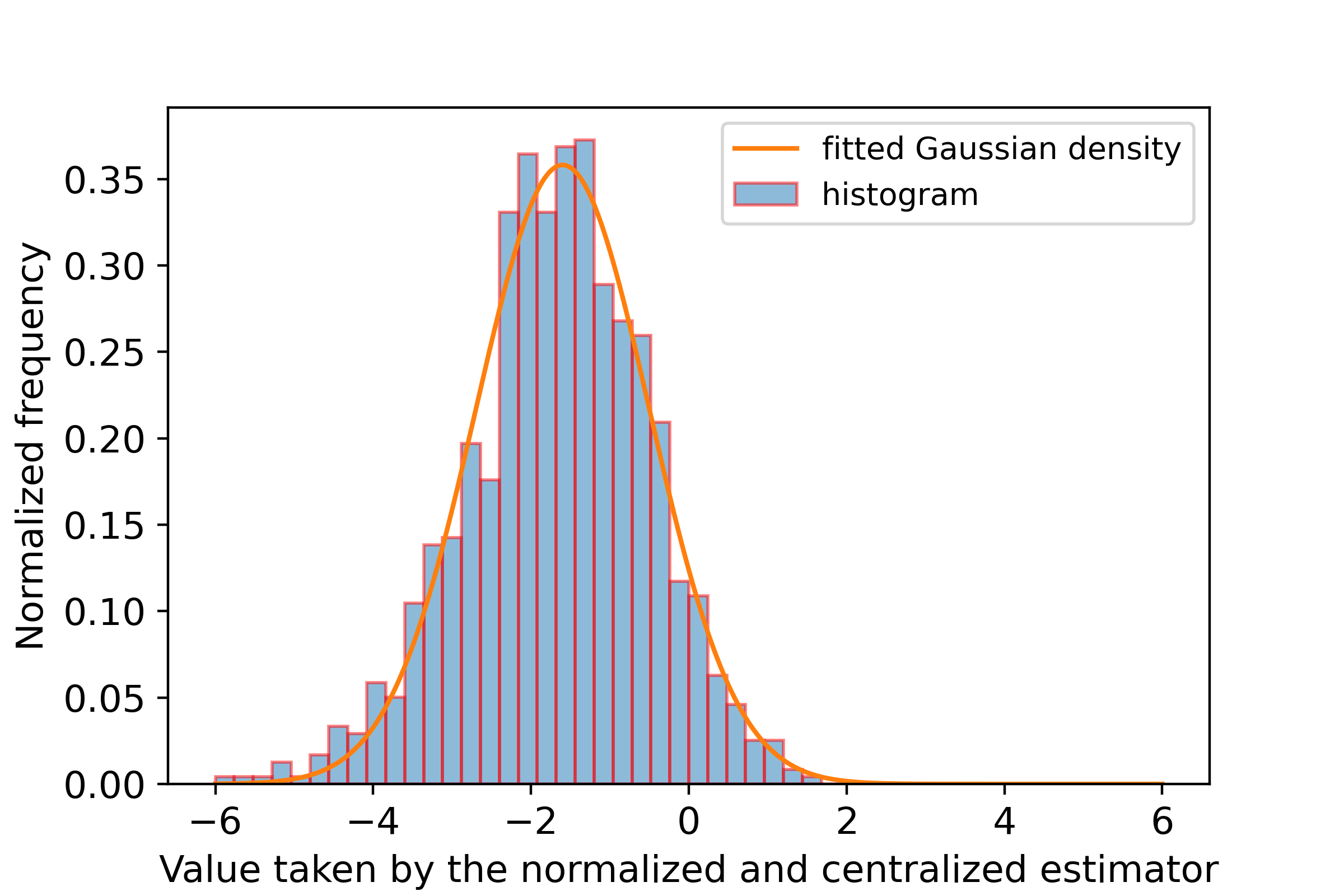}}
 \caption{}
 \end{figure}

\begin{figure}[h]
 \centering 
 \subfigure[Estimator distribution based on  Algorithm \ref{alg:xxx}, $T=10^6$]{
 \label{fig:1e6_algo}
 \includegraphics[scale=0.44]{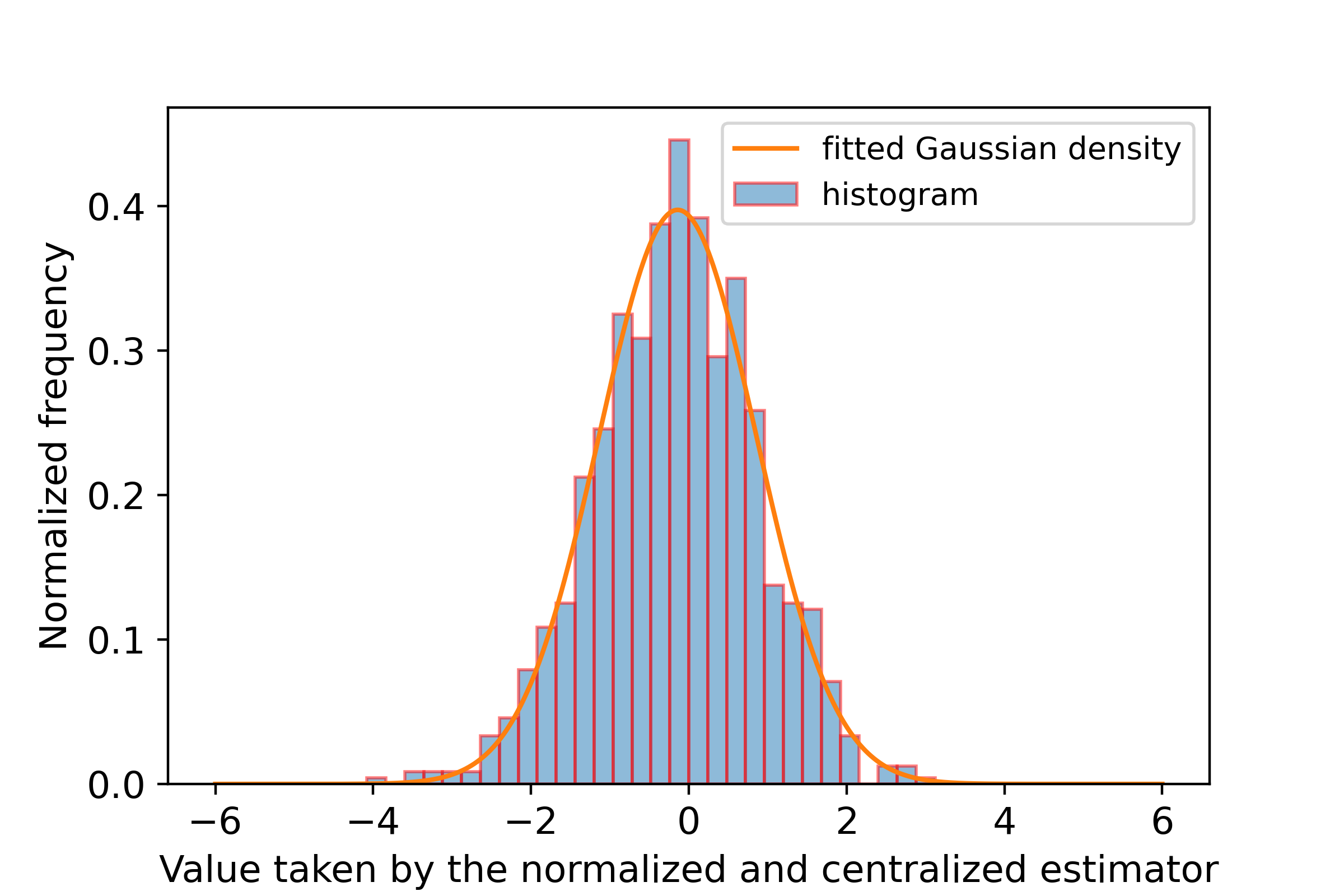}}
 \subfigure[Estimator distribution based on  two-point central FD method, $T=10^6$]{
  \label{fig:1e6_naive}
 \includegraphics[scale=0.44]{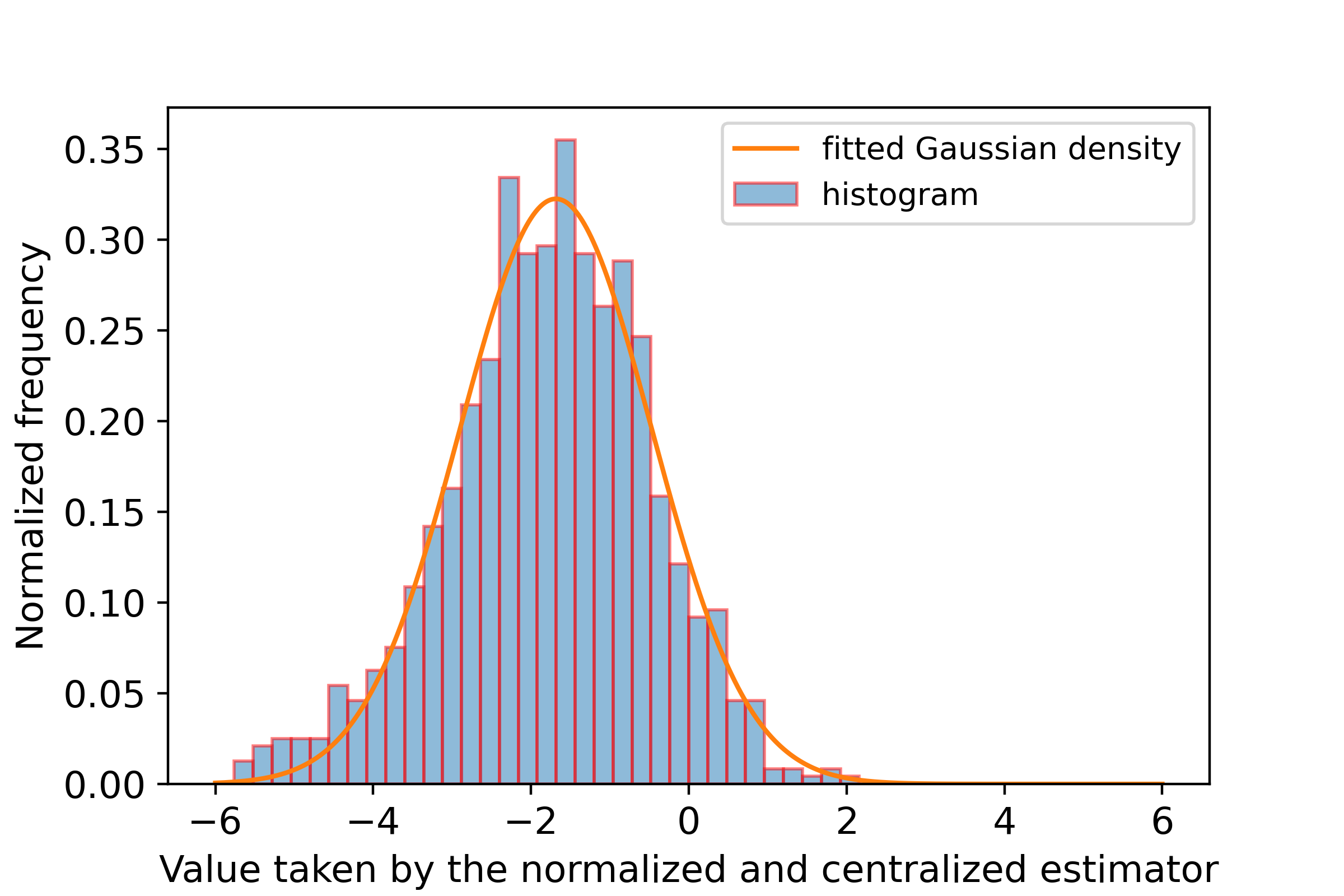}}
  \caption{}
 \end{figure}

\begin{figure}[h]
 \centering 
 \subfigure[Estimator distribution based on  Algorithm \ref{alg:xxx}, $T=10^7$]{
 \label{fig:1e7_algo}
 \includegraphics[scale=0.44]{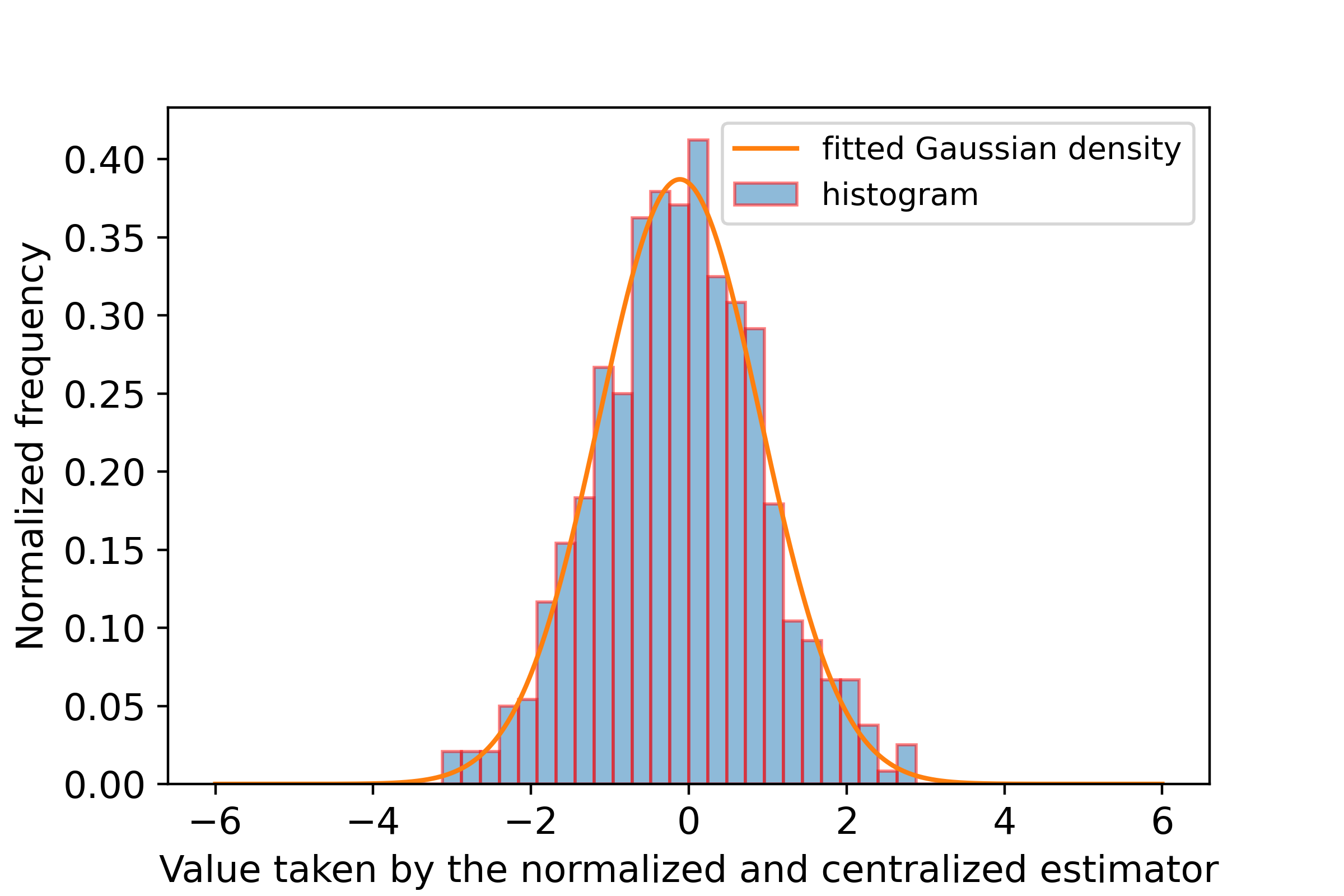}}
 \subfigure[Estimator distribution based on two-point central FD method, $T=10^7$]{
  \label{fig:1e7_naive}
 \includegraphics[scale=0.44]{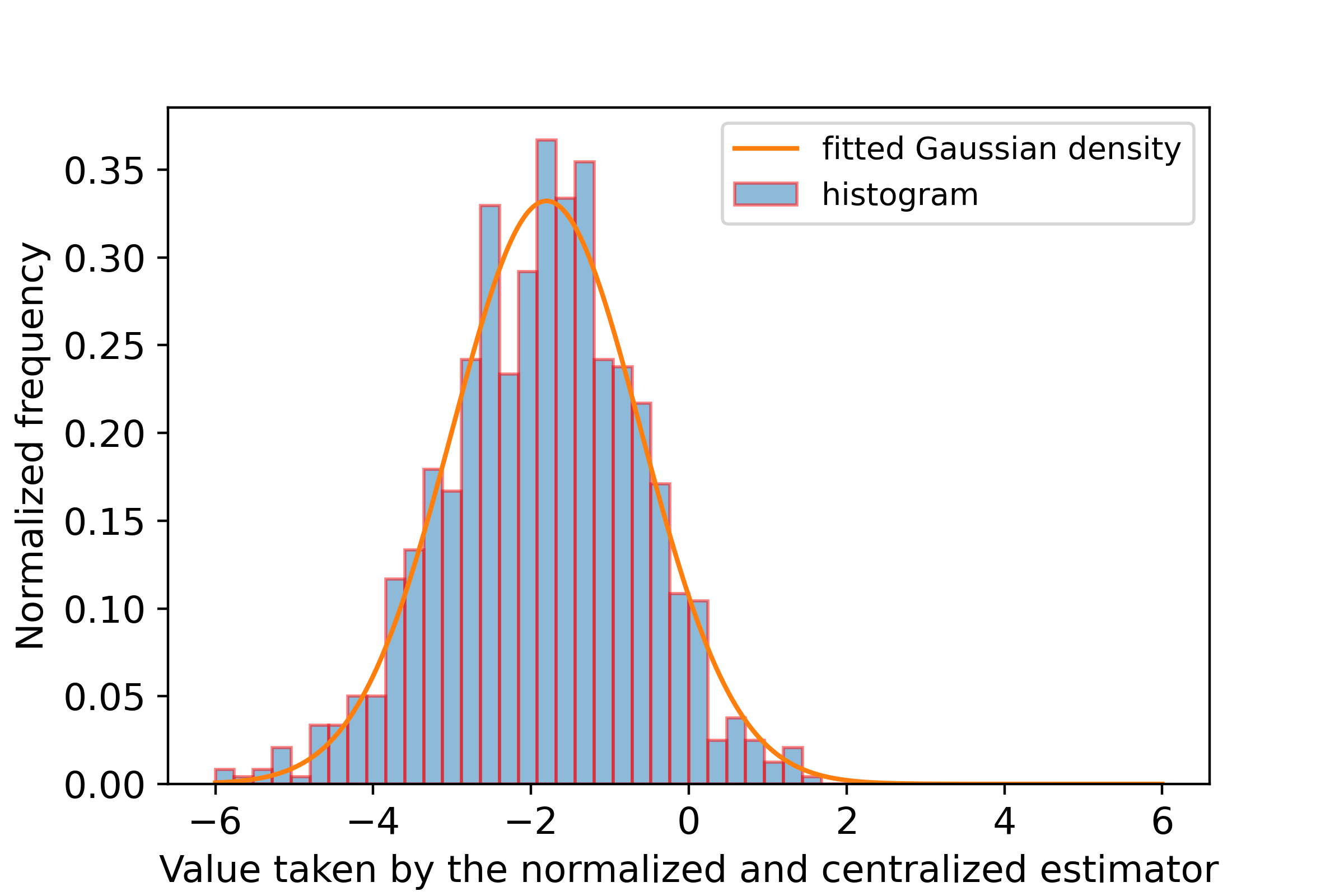}}
  \caption{}
 \end{figure}

If our algorithm is not used but the standard two-point central finite difference (FD) method is used to construct gradient estimator and function value estimator, Figure \ref{fig:1e5_naive}, \ref{fig:1e6_naive} and \ref{fig:1e7_naive} show the distribution of the normalized and centralized estimator given by central FD method with sample size $T=10^5,10^6,10^7$, respectively. From those results in Figure \ref{fig:1e5_naive}, \ref{fig:1e6_naive} and \ref{fig:1e7_naive} , we can see that the estimators provided by the standard two-point central FD method have a significant non-zero bias (because we know the true underlying optimal value) and are centered around $-2$ instead of $0$. Because such bias is generally impossible to accurately estimate without knowing the true optimal value, it is not reliable to use the corresponding estimator to provide valid confidence intervals. For example, we would not know how to accurately decide the center of the confidence interval due to the unknown and non-vanishing bias.

In comparison, Figure \ref{fig:1e5_algo}, \ref{fig:1e6_algo} and \ref{fig:1e7_algo} show that the estimators given by our Algorithm \ref{alg:xxx} centered closely to $0$ and are nearly normally distributed. In addition, the coverage rates of the $95\%$ confidence interval (CI) we constructed are $93.5\%,93.9\%$ and $94.2\%$ for $T=10^5,10^6$ and $10^7$, respectively, so valid statistical inference is possible for our estimator.

\begin{figure}[]
  \centering 
  \subfigure[Estimator distribution based on Algorithm \ref{alg:xxx}, Pareto distribution]{
  \label{fig:pareto_algo}
  \includegraphics[scale=0.44]{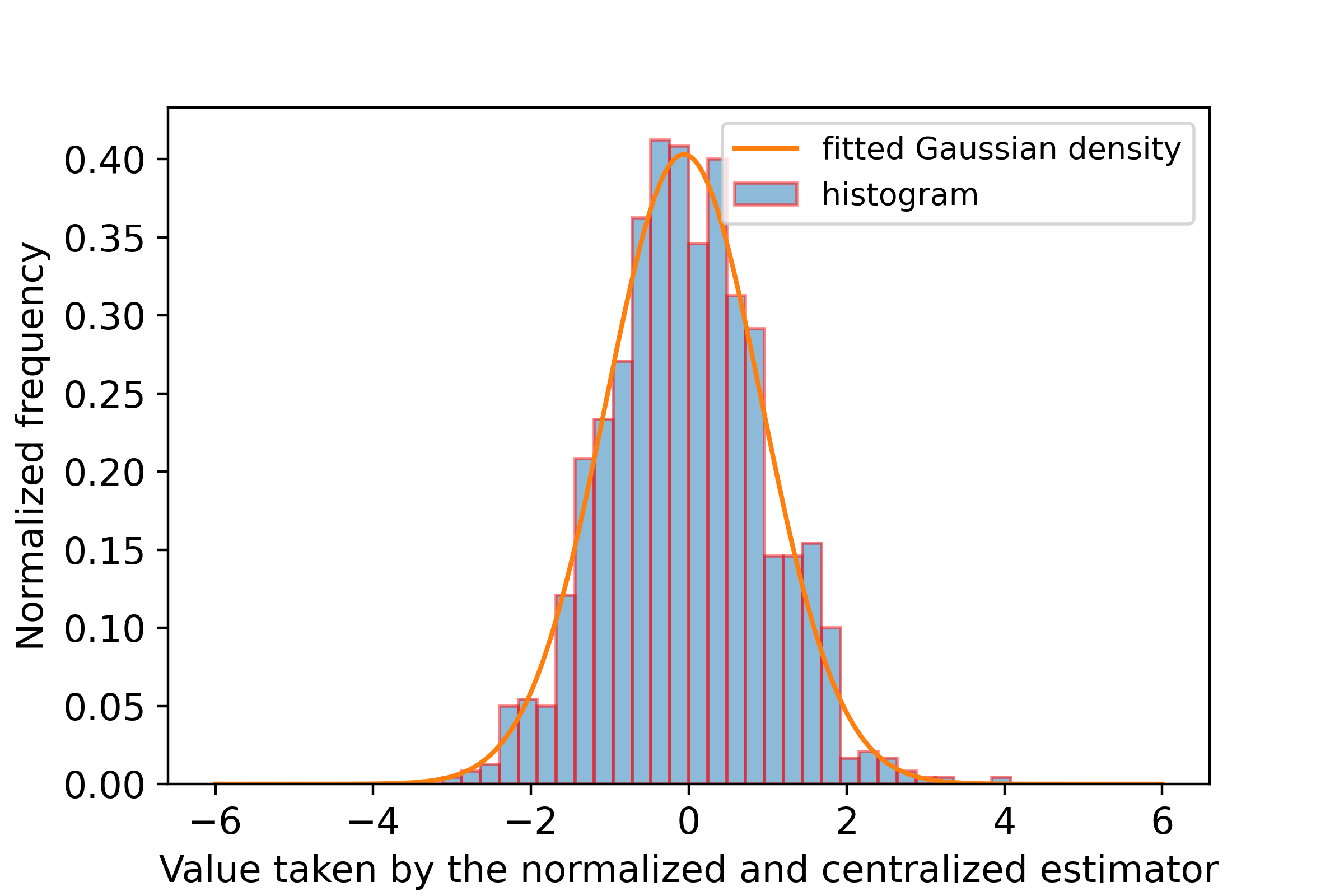}}
  \subfigure[Estimator distribution based on  two-point central FD method, Pareto distribution]{
  \label{fig:pareto_naive}
  \includegraphics[scale=0.44]{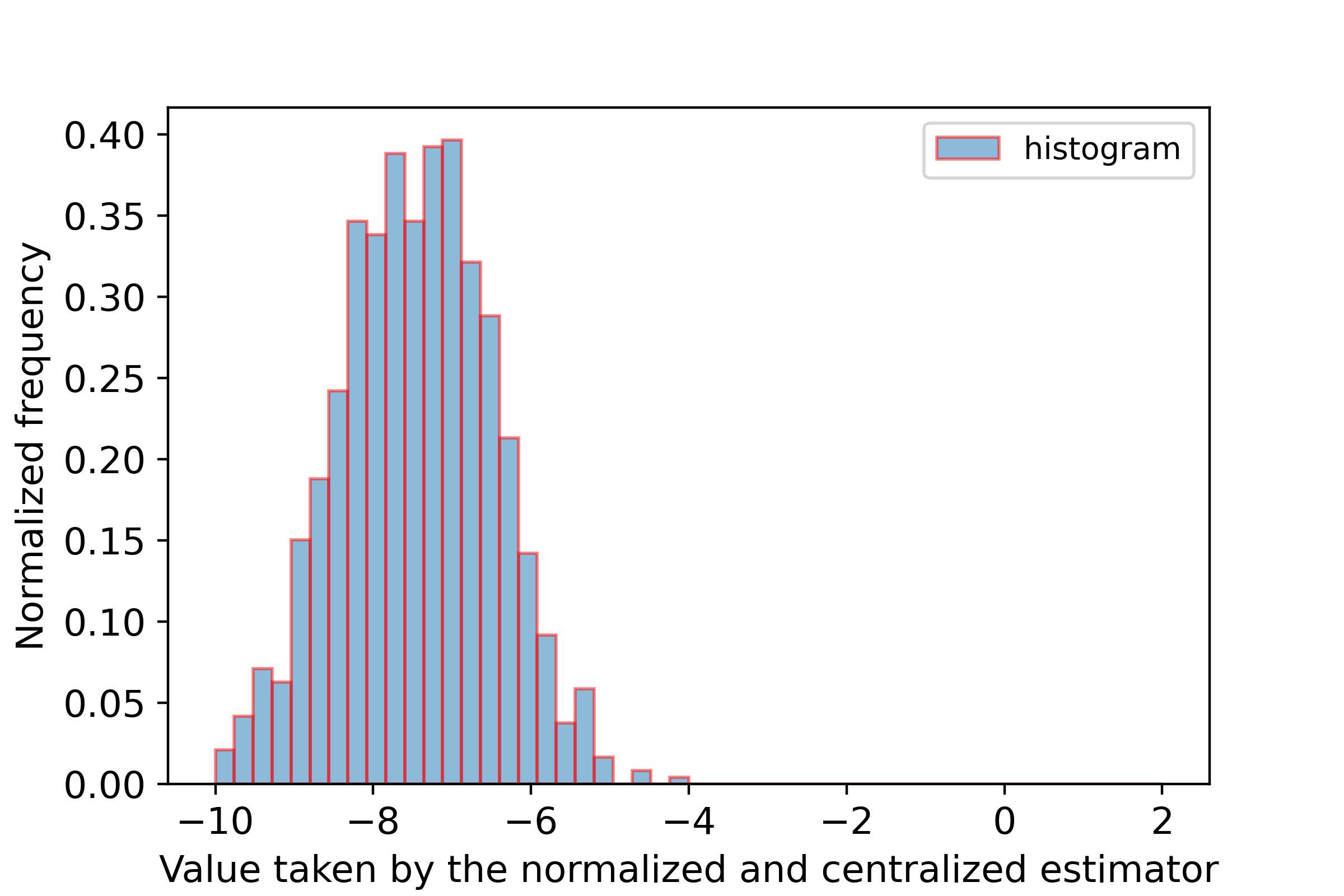}}
   \caption{}
  \end{figure}

 Next we examine the performance of our algorithm under other outcome distributions. We use the same parameters and the same $\mu$, but this time $Y(\theta)$ is a random variable with Pareto distribution, which is known to be heavy-tailed. The shape parameter of this distribution is chosen to be $\alpha=3$, and the scale parameter is chosen to be $\frac{2\mu(\theta)}{3}$, so $\E Y(\theta)=\mu(\theta)$. The results are given by Figure \ref{fig:pareto_algo} and \ref{fig:pareto_naive}.
 
 This time, the estimator provided by our algorithm still works well and the coverage rate of $95\%$ CI is $94.5\%$, while the central FD method gives an estimator with bias as large as $-8$ and completely fails. In addition to the Pareto distribution, we also consider the case that the mean of $Y(\theta)$ is $\mu(\theta)$ but the noise is given by a standard t distribution with degree of freedom is $3$. The results are given in the Appendix. 
 
\begin{figure}[h]
 \centering 
 \label{fig:3}
 \includegraphics[scale=0.5]{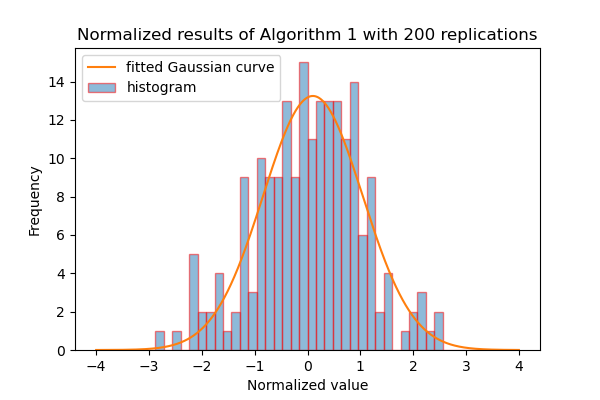}
 \caption{Result of Algorithm \ref{alg:xxx}, multi-dimensional case}
 \end{figure}

Finally, we provide the results of a multi-dimensional synthetic numerical experiment. In this experiment we set $d=6$ and $\theta=(\theta_1,\cdots,\theta_6)^T\in[0,1]^6$. $\mu(\theta)$ is given by following logistic model:
$$\mu(\theta)=-\frac{\exp(-\frac{1}{2}\sum_{i=1}^d(\theta_i-\frac{1}{3})^2-2)}{1+\exp(-\frac{1}{2}\sum_{i=1}^d(\theta_i-\frac{1}{3})^2-2)}.$$ Again we assume that $-\mu(\theta)$ denotes the CTR, and set $k=3, m=100,\nu=\frac{1}{5},\theta^0=(\frac{1}{2},\cdots,\frac{1}{2})^T,T=10^7$, and $c_t=\frac{1}{t^\nu}$, $\alpha_t=\frac{20}{t}$, respectively. We repeat our algorithm $200$ times and report the normalized estimator as Figure \ref{fig:3}. The estimator is nearly normal distributed. In addition, the $95\%$ CI has a $94.0\%$ coverage rate.

In this section, the numerical experiments' implication is that, classical zeroth-order stochastic optimization algorithms using two-point finite difference gradient estimator, despite of its convergence rate guarantee on the optimizer, may fail to provide a valid unbiased central limit theorem for the purpose of valid statistical inference. Our algorithm is proved to fix this issue and is illustrated by the above synthetic experiments.

\section{Conclusion}\label{sec:conclusion}
We consider dual tasks of optimization and statistical inference for a class of simulation optimization tasks. We find that classical optimization algorithms, despite of their fast convergence rates under convexity assumptions, do not come with a central limit theorem with vanishing bias that can be used to construct asymptotically valid confidence intervals.  We provide  a new optimization algorithm that on one hand maintains the same fast convergence rate under convexity assumptions and on the other hand permits the establishment of a valid central limit theorem.

\bibliographystyle{informs2014} 
\bibliography{ref}


\newpage

\section*{Appendix} \label{appendix}


\, 

\subsection*{Proof for Section 3}

{\noindent\bf Proof of Proposition \ref{prop:var_m}.} 
$$
\begin{aligned}
T\Var(\hat{\mu}(\hat{\theta}_T))=\frac{4m^2}{T}\sum_{i=1}^{T/(2m)}\Var(\hat{\mu}_i).\\
\end{aligned}
$$
For each $i$, note that we have:
$$\Var(\hat{\mu}_i)=\frac{\frac{\sigma(\theta^{++})^2}{m_2}+\frac{k^4\sigma(\theta^+)^2}{m_1}+\frac{k^4\sigma(\theta^-)^2}{m_1}+\frac{\sigma(\theta^{--})^2}{m_2}}{4(k^2-1)^2}.$$
By Theorem \ref{thm:general_error}, $\theta^t\to\theta^*$ with probability $1$, then by the continuity of $\sigma(\cdot)$, we know $\Var(\hat{\mu}_i)\to\frac{\sigma(\theta^*)^2\big(\frac{1}{m_2}+\frac{k^4}{m_1}\big)}{2(k^2-1)^2}.$ Now, plug in and we obtain
$$T\Var(\hat{\mu}(\hat{\theta}_T))\to \frac{m\sigma(\theta^*)^2\big(\frac{1}{m_2}+\frac{k^4}{m_1}\big)}{(k^2-1)^2},$$
with probability $1$.

\subsection*{Proofs for Section 4}
{\noindent\bf Proof of Theorem \ref{thm:general_error}.}

We will follow the idea of Theorem 3.1 of \cite{l1998budget}. Note that in the proof of Theorem 3.1 of \cite{l1998budget}, they consider the case $\rho=1$ and $\rho<1$ separately and show that they are similar. In our case, we also prove the case $\rho=1$ and the same procedure can be applied to the case that $\rho<1$. Suppose we now want to minimize $\eta(\theta)$. Then the updating procedure of $\theta_t$ is now $$
\theta^{t+1}=\theta^t-\alpha_tg^t.$$
Note that we can write $g^t$ as:$$g^t=\big(\nabla \eta(\theta^t)^Tw^t\big)w^t+b_t+(g^t-\E[ g^t\big|w^t,\theta^t]),$$ here $b_t$ denotes the error of finite difference scheme, the last term is the random error comes from the sampling of $y_i$, and the $\big(\nabla \eta(\theta^t)^Tw^t\big)$ is the directional derivative at $\theta^t$. 

WLOG we assume $\theta^*=0$, then
$$
\begin{aligned}
&\E\Vert\theta^{t+1}\Vert_2^2-\E\Vert\theta^t\Vert_2^2\\
=&\E\Vert\theta^{t+1}-\theta^t\Vert_2^2+2\E(\theta^{t+1}-\theta^t)^T(\theta^t)\\
=&\E\Vert\theta^{t+1}-\theta^t\Vert_2^2-2\alpha_t\E (g^t)^T(\theta^t).
\end{aligned}
$$
Since we have
$$\theta^{t+1}-\theta^t=-\alpha_tg^t,$$
and by Taylor expansion we have:
$$\nabla\eta(\theta^t)=\nabla^2\eta(\theta^*)\theta^t+O(\Vert\theta^t\Vert_2^2),$$
plug in and we obtain:
$$
\begin{aligned}
&\E\Vert\theta^{t+1}\Vert_2^2-\E\Vert\theta^t\Vert_2^2\\
=&\E\Vert\theta^{t+1}-\theta^t\Vert_2^2-2\alpha_t\E b_t^T\theta^t-2\alpha_t\E\big(\nabla\eta(\theta^t)^Tw^t\big)(w^t)^T(\theta^t)\\
=&\alpha_t^2\E\Vert g^t\Vert_2^2-2\alpha_t\E b_t^T\theta^t-2\alpha_t\E\big((\theta^t)^T\nabla^2\eta(\theta^t)w^t(w^t)^T\theta^t\big)\\
&+\alpha_tO(\E\Vert\theta^t\Vert_2^3).
\end{aligned}$$
Note that
$$
\begin{aligned}
&\E(\theta^t)^T\nabla^2\eta(\theta^t)w^t(w^t)^T\theta^t\\
=&\E \text{tr}\big(w^t(w^t)^T\theta^t(\theta^t)^T\nabla^2\eta(\theta^t)\big)\\
=&\text{tr}\big(\E[w^t(w^t)^T]\theta^t(\theta^t)^T\nabla^2\eta(\theta^t)\big)\\
=&\text{tr}\big(\frac{I_d}{d}\E\theta^t(\theta^t)^T\nabla^2\eta(\theta^t)\big)\\
\geq&\frac{\lambda_\text{min}}{d}\E\Vert\theta^t\Vert_2^2,
\end{aligned}$$
here $d$ is the dimension of $\theta$ and $\lambda_\text{min}$ is the minimum eigenvalue of $\nabla^2\mu(\theta^*)$ and is positive since we assume that $\eta(\cdot)$ is strongly convex. Then we have:
\begin{equation}\label{proof:new_d}
\begin{aligned}
&\E\Vert\theta^{t+1}\Vert_2^2-\E\Vert\theta^t\Vert_2^2\\
\leq& -2\frac{c_0}{dt}\lambda_\text{min}\E\Vert\theta^t\Vert_2^2+\frac{c_0}{t}O(\E\Vert\theta^t\Vert_2^3)-\frac{2c_0}{t}\E b_t^T\theta^t\\
&+\frac{3c_0^2}{t^2}\Big(\E(\nabla\eta(\theta^t)^Tw^t)^2+\E \Vert b_t\Vert^2+\Var(g^t\big|w^t,\theta^t)\Big).
\end{aligned}
\end{equation}

Now, if we replace $(\text{A}.3)$ in the proof of Theorem 3.1 in \cite{l1998budget} by the inequality above, then the proof can be  adapted from the proof of Theorem 3.1 in \cite{l1998budget}. This proof is given as follows.

By the continuity of $\nabla\eta(\theta)$, we can find a constant $K_\alpha>0$ s.t. $$\E(\nabla\eta(\theta^t)^Tw^t)^2\leq \E \Vert \nabla \eta(\theta^t)\Vert^2\leq 2K_\alpha(1+\E\Vert\theta^t \Vert^2).$$
We also have 
$$\E \Vert b_t^T\theta^t\Vert\leq \E^{1/2}\Vert b_t\Vert^2\E^{1/2}\Vert\theta^t\Vert^2.$$
By the property of central FD method, there exists $K_\beta>0$ s.t.
$$\E^{1/2}\Vert b_t\Vert^2\leq K_\beta t^{-2\nu},$$
then by basic inequality, we have:
$$
\begin{aligned}
&\E \Vert b_t^T\theta^t\Vert \\
\leq&K_\beta t^{-2\nu}\big(t^{2\nu}\lambda_\text{min}/(2dK_\beta)\E\Vert\theta^t\Vert^2+K_\beta d/(2\lambda_\text{min})t^{-2\nu}\big),
\end{aligned}$$
which then gives:
$$\E \Vert b_t^T\theta^t\Vert \leq\frac{\lambda_\text{min}}{2d}\E\Vert\theta^t\Vert^2+K_\beta^2d/(2\lambda_\text{min})t^{-4\nu}.$$
For the $O(\E\Vert\theta^t\Vert_2^3)$ term in \eqref{proof:new_d}, since $\theta$ is bounded, it can easily be controlled by $K_1(1+\E\Vert \theta^t\Vert^2)$ with some $K_1>0$.
Now, plug above results in to \eqref{proof:new_d} and we obtain:
\begin{equation}\label{proof:new_d_2}
\begin{aligned}
&\E\Vert\theta^{t+1}\Vert_2^2-\E\Vert\theta^t\Vert_2^2\\
\leq& -\frac{c_0}{dt}\lambda_\text{min}\E\Vert\theta^t\Vert_2^2+\frac{c_0K_1}{t}+\frac{c_0K_1}{t}\E\Vert \theta^t\Vert^2\\&+{c_0}K_\beta^2d/(\lambda_\text{min})t^{-4\nu-1}
+\frac{6c_0^2}{t^2}K_\alpha(1+\E\Vert\theta^t \Vert^2)\\&+{3c_0^2K_\beta^2}t^{-4\nu-2}+{3c_0^2}K_\delta t^{2\nu-2}.
\end{aligned}
\end{equation}
Now, take $0<\lambda_0<c_0\lambda_\text{min}/d$, rearrange terms and take $t\geq t_0$ large enough we have:
\begin{equation}\label{analog_A8}
\begin{aligned}
&\E\Vert\theta^{t+1}\Vert_2^2\\
\leq& (1-\frac{\lambda_0}{t})\E\Vert\theta^t\Vert_2^2+\frac{2c_0K_1}{t}\E\Vert \theta^t\Vert^2+\frac{2{c_0}K_\beta^2d}{\lambda_\text{min}}t^{-4\nu-1}\\&+{3c_0^2}K_\delta t^{2\nu-2}+\frac{K_2}{t},
\end{aligned}
\end{equation}
here $K_2>0$ is some constant. Note that, this result \eqref{analog_A8} adapts similar form with $(\text{A}.8)$ in \cite{l1998budget} (their parameters $\gamma$, $\beta$, $\delta$ correspond to our $1$, $2\nu$ and $-2\nu$). Apply \eqref{analog_A8} iteratively and we obtain:$$
\begin{aligned}
\E\Vert\theta^{t+1}\Vert_2^2\leq& A_{tt_0}\E\Vert\theta^{t_0}\Vert_2^2+\sum_{i=t_0}^t\frac{2c_0K_1}{i}A_{ti}\E\Vert \theta^i\Vert^2\\ &+\frac{2{c_0}K_\beta^2d}{\lambda_\text{min}}\sum_{i=t_0}^tA_{ti}i^{-4\nu-1}\\
&+{3c_0^2}K_\delta \sum_{i=t_0}^tA_{ti}i^{2\nu-2}+\sum_{i=t_0}^tA_{ti}\frac{K_2}{i},
\end{aligned}$$
here $A_{tj} = \prod_{k=j+1}^t(1-\lambda_0k^{-1})$ if $j<t$, otherwise $A_{tj}=1$.


 Since we are only interested in the order of $\E\Vert \theta^t\Vert^2$, a difference in constant factors will actually not affect our results. Hence, we can calculate the order of those summations of $A_{tj}$ exactly the same as follows: for each fixed $j<t$,
$$
\begin{aligned}|A_{tj}|\leq &\exp\big(-\lambda_0\sum_{k=j+1}^tk^{-1}\big)\leq \exp(\lambda_0/t-\lambda_0\int_{j}^tx^{-1}dx)\\
&=\exp(\lambda_0/j)(j/t)^{\lambda_0}.
\end{aligned}$$
We thus have
$$
\begin{aligned}
&\sum_{i=t_0}^t|A_{ti}|i^{-4\nu-1}\leq\exp(\lambda_0/t_0)\sum_{i=t_0}^{t-1}i^{-4\nu-1}(i/t)^{\lambda_0}\\
&\leq \exp(\lambda_0/t_0)(\lambda_0-4\nu)^{-1}t^{-4\nu}.
\end{aligned}$$
Similarly,
$$\sum_{i=t_0}^{t-1}i^{2\nu-2}|A_{ti}|\leq\exp(\lambda_0/t_0)(\lambda_0+2\nu-1)^{-1}t^{2\nu-1}.$$
In addition, since $$A_{tj}-A_{t,j-1}=\lambda_0j^{-1}A_{tj},$$  we have
$$\sum_{i=t_0+1}^{t}A_{ti}i^{-1}=A_{tt}-A_{t,t_0}.$$
Thus, there exist some $K_3>0$ and some $\epsilon_0>0$:
$$
\begin{aligned}
\E\Vert\theta^{t+1}\Vert_2^2\leq& \sum_{i=t_0}^t\frac{2c_0K_1}{i}A_{ti}\E\Vert \theta^i\Vert^2+\frac{2c_0K_\beta^2d}{\lambda_\text{min}\lambda_0(1-\epsilon_0/\lambda_0)}t^{-4\nu}\\
&+\frac{3c_0^2K_\delta}{\lambda_0(1-\epsilon_0/\lambda_0)}t^{2\nu-1}+K_3+O(t^{-4\nu-1}+t^{-2+2\nu}).
\end{aligned}
$$
Now, by Gronwall's inequality we know that $\sup_t\Vert\theta^t\Vert^2<K_4$ for some $K_4>0$. By Assumption \ref{assumption:ode} and Theorem 2.3.1 of \cite{kushner2012stochastic}, we know $\theta^t$ converges to $\theta^*$ w.p.1.. Combine these two results together we obtain that $\E \Vert\theta^t\Vert^2\to0=\E\Vert \theta^*\Vert^2$ by our assumption.

Now, to obtain the final result, note that
 this time we will not replace the $O(\E\Vert\theta^t\Vert_2^3)$ term in \eqref{proof:new_d} by $K_1(1+\E\Vert \theta^t\Vert^2)$, but bound it by $\frac{\lambda_0}{2c_0}\E\Vert \theta^t\Vert^2$, which is possible for large enough $t$ based on the previous result $\E\Vert \theta^t\Vert^2\to 0$. After that, we proceed exactly the same as before, then $\sum_{i=t_0}^tA_{it}\E\Vert\theta^i\Vert^2$ and $\sum_{i=t_0}^tA_{ti}i^{-1}$ terms will not appear, thus the inequality above will become:
$$
\begin{aligned}
&\E\Vert\theta^{t+1}\Vert_2^2\leq \frac{4c_0K_\beta^2d}{\lambda_\text{min}\lambda_0(1-2\epsilon_0/\lambda_0)}t^{-4\nu}
+\frac{6c_0^2K_\delta}{\lambda_0(1-2\epsilon_0/\lambda_0)}t^{2\nu-1}\\
&+O(t^{-4\nu-1}+t^{-2+2\nu}).
\end{aligned}
$$
Since $\frac{1}{6}<\nu<\frac{1}{4}$, then $-1+2\nu>-4\nu$, so $\E\Vert\theta^{t}\Vert_2^2=O(t^{-1+2\nu})$, which finishes the proof.
\color{black}

\,

{\noindent\bf Proof of Theorem \ref{thm:clt_simple}.}
Since $\mu(\cdot)$ is third-order differentiable, we have 
$$\mu(\theta+h)+\mu(\theta-h)=2\mu(\theta)+h^T\nabla^2\mu(\theta)h+O(\Vert h\Vert_2^3),$$
then plug in and we obtain:
$$\mu(\theta^+)+\mu(\theta^-)=2\mu(\theta)+c_t^2(w^t)^T\nabla^2w^t+O(c_t^3),$$
and
$$\mu(\theta^{++})+\mu(\theta^{--})=2\mu(\theta)+k^2c_t^2(w^t)^T\nabla^2w^t+O(c_t^3).$$
From above two equations we have: \begin{equation}\label{proof:val_error}
{\mu}(\theta^t)=\frac{-{\mu}(\theta^{++})+k^2{\mu}(\theta^{+})+k^2\mu(\theta^-)-\mu(\theta^{--})}{2(k^2-1)}+O(c_t^3).\end{equation}

Now, if we decompose $\hat{\mu}_t$ in the following way:
$$
\begin{aligned}
\hat{\mu}_t&=\big(\hat{\mu}_t-\E[\hat{\mu}_t\big|\theta^{t-1}]\big)\\&+\big(\E[\hat{\mu}_t\big|\theta^{t-1}]-\E[\mu(\theta^t)\big|\theta^{t-1}]\big)+\E[\mu(\theta^t)\big|\theta^{t-1}],
\end{aligned}$$
then by \eqref{proof:val_error}, we have
$$\hat{\mu}_t=\big(\hat{\mu}_t-\E[\hat{\mu}_t\big|\theta^{t-1}]\big)+\E[\mu(\theta^t)\big|\theta^{t-1}]+O_p(c_t^3).$$

Then $$
\begin{aligned}
&\sqrt{T}\big(\frac{1}{T}\sum_{i=1}^{T/(2m)}\hat\mu_i-\mu(\theta^*)\big)\\
&=\frac{1}{\sqrt{T}}\sum_{i=1}^{T/(2m)}\big(\hat{\mu}_t-\E[\hat{\mu}_t\big|\theta^{t-1}]\big)\\
&+\frac{1}{\sqrt{T}}\big(\sum_{i=1}^{T/(2m)}(\E[\mu(\theta^t)\big|\theta^{t-1}]-\mu(\theta^*))\big)+O_p(T^{\frac{1}{2}-3\nu}).
\end{aligned}$$
As for the first term, from Assumption \ref{assumption:noise} we know Lindeberg condition is satisfied, so by martingale CLT we obtain a CLT for the first summation. By Proposition \ref{prop:var_m} we know the asymptotic variance is $\frac{(k^2+1)^2\sigma(\theta^*)^2}{(k^2-1)^2}$ if we set $m_1,m_2$ as we discussed in Section \ref{subsec:choice_of_para}. The third term is $o_p(1)$ since $\nu>\frac{1}{6}$. Now we only need to control the second term, note that 
$$
\begin{aligned}
&\E \big|\sum_{i=1}^{T/(2m)}(\E[\mu(\theta^t)\big|\theta^{t-1}]-\mu(\theta^*))\big|\\
\leq &\E\sum_{i=1}^{T/(2m)}\E[|\mu(\theta^t)-\mu(\theta^*)|\big|\theta^{t-1}]\\
= & \sum_{i=1}^{T/(2m)}\E[|\mu(\theta^t)-\mu(\theta^*)|]\\
\leq &M\sum_{i=1}^{T/(2m)}\E\Vert\theta^t-\theta^*\Vert_2^2.
\end{aligned}$$
Here $M=\sup_{\theta\in[0,1]^d}\lambda_\text{max}\big(\nabla^2\mu(\theta)\big)$, since $\nabla^2\mu$ is continuous, so $M<+\infty$. From Theorem \ref{thm:general_error} we know $\E\Vert\theta^t-\theta^*\Vert_2^2=O(t^{-1+2\nu})$, so the second term is $O_p(T^{-\frac{1}{2}+2\nu})=o_p(1)$. Combine all above results together we obtain the following CLT:
$$
\sqrt{T}(\hat{\mu}(\hat{\theta}_T) - \mu(\theta^*)) \overset{d.}{\rightarrow} \frac{(k^2+1)\sigma(\theta^*)}{k^2-1} \mathcal{N}(0,1) 
$$
as $T\to+\infty$. Similarly we have
$$\sqrt{T}\E|\mu(\hat\theta_T)-\mu(\theta^*)|=o(1),$$ so $\sqrt{T}\big(\mu(\hat\theta_T)-\mu(\theta^*)\big)=o_p(1)$, which then gives $$
\sqrt{T}(\hat{\mu}(\hat{\theta}_T) - \mu(\hat\theta_T)) \overset{d.}{\rightarrow} \frac{(k^2+1)\sigma(\theta^*)}{k^2-1} \mathcal{N}(0,1) 
$$
as $T\to+\infty$. 

{\noindent\bf Consistency of $\hat{\sigma}(\theta^*)$.}
We will need an extra condition that the fourth order moment of $\epsilon_\theta$ are uniformly bounded for all $\theta\in[0,1]^d$.
The estimator of $\sigma(\theta^*)$ is given by: (we take $I=1$ in our algorithm)
$$\hat{\sigma}^2(\theta^*)=\sum_{i=I}^T(y_i-\bar{y})^2/(T-I+1),$$
here $\bar{y}=\frac{1}{T-I+1}\sum_{i=I}^Ty_i$. Recall that $y_i$ is a realization of $Y(\theta_i)$, we have
$$\hat{\sigma}^2(\theta^*)=\frac{1}{T-I+1}\sum_{i=I}^T(\mu(\theta_i)+\sigma(\theta_i)\cdot\epsilon_{\theta_i}-\bar{y})^2.$$
Expand the square and we have:
$$
\begin{aligned}
\hat{\sigma}^2(\theta^*)&=\frac{1}{T-I+1}\sum_{i=I}^T(\mu(\theta_i)-\bar{\mu})^2\\
&+\frac{1}{T-I+1}\sum_{i=I}^T(\sigma(\theta_i)\cdot\epsilon_{\theta_i}-\bar{\sigma})^2+O_p(\frac{1}{T}),
\end{aligned}$$
here $\bar\mu=\frac{1}{T-I+1}\sum_{i=I}^T\mu(\theta_i)$ and $\bar{\sigma}=\frac{1}{T-I+1}\sum_{i=I}^T\sigma(\theta_i)\cdot\epsilon_{\theta_i}$. Then
$$\begin{aligned}
&\hat{\sigma}^2(\theta^*)=\frac{1}{T-I+1}\sum_{i=I}^T\mu^2(\theta_i)-\bar{\mu}^2\\
&+\frac{1}{T-I+1}\sum_{i=I}^T\sigma^2(\theta_i)\cdot\epsilon^2_{\theta_i}-\bar{\sigma}^2+O_p(\frac{1}{T}).
\end{aligned}$$
For the first two terms, since $\mu(\theta_i)\to\mu(\theta)^*$ and $\bar{\mu}\to\mu(\theta)^* $, by Stolz theorem we know it's $o(1)$. For $\bar{\sigma}^2$, by CLT we know it's $O_p(\frac{1}{T})$. As for $\frac{1}{T-I+1}\sum_{i=I}^T\sigma^2(\theta_i)\cdot\epsilon^2_{\theta_i}$, by our extra condition it is $\frac{1}{T-I+1}\sum_{i=I}^T\sigma^2(\theta_i)$. Since $\sigma(\theta_i)\to \sigma(\theta^*)$, again by Stolz theorem we  know $\frac{1}{T-I+1}\sum_{i=I}^T\sigma^2(\theta_i)=\sigma^2(\theta^*)+o(1)$. Thus, $\hat{\sigma}^2(\theta^*)=\sigma^2(\theta^*)+o_p(1)$, which finishes our proof.

\,

\begin{figure}[H]
 \centering 
 \subfigure[Result of Algorithm \ref{alg:xxx}, t distribution]{
   \label{fig:t_algo}
 \includegraphics[scale=0.44]{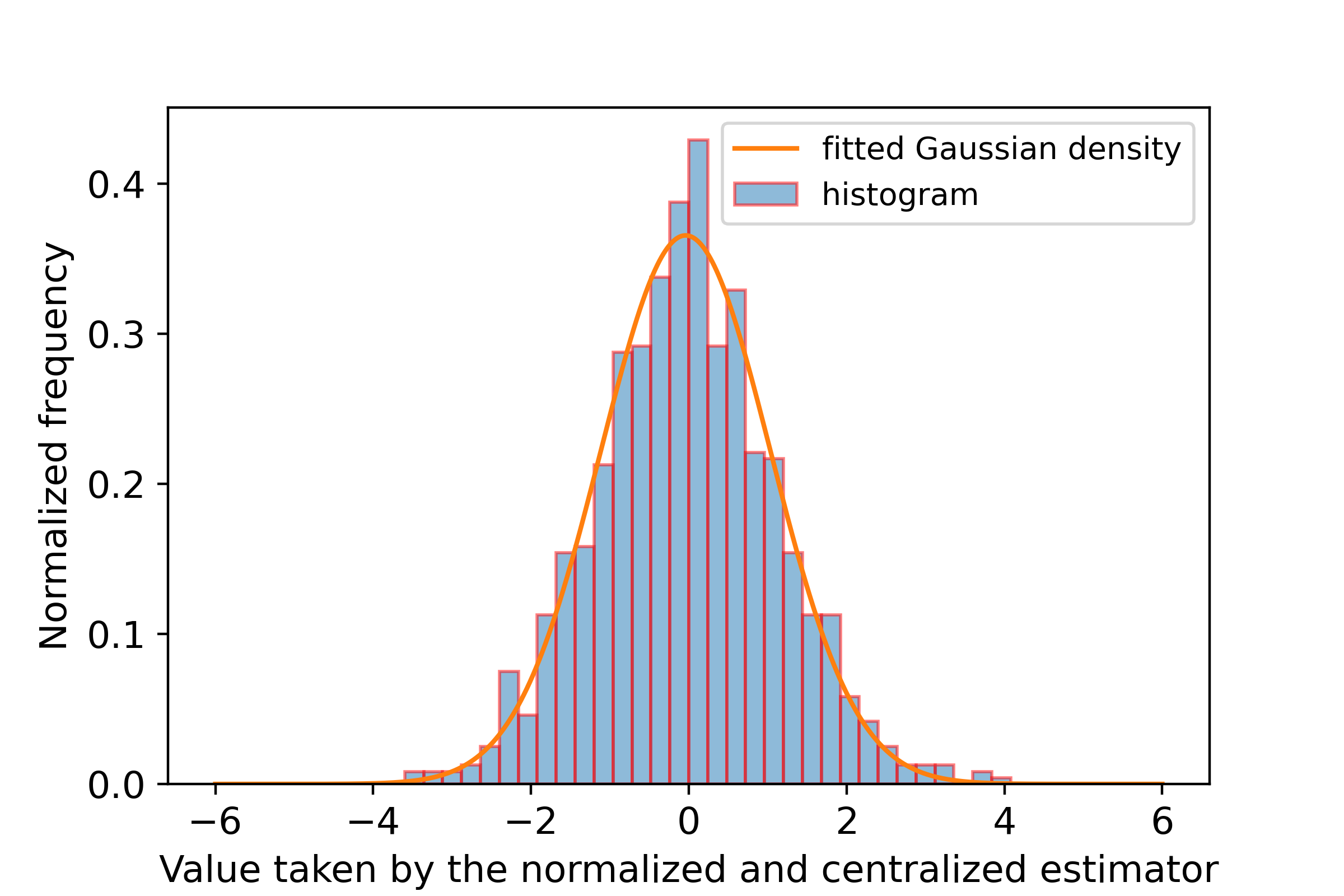}}
 \subfigure[Result of two-point central FD method, t distribution]{
   \label{fig:t_naive}
 \includegraphics[scale=0.44]{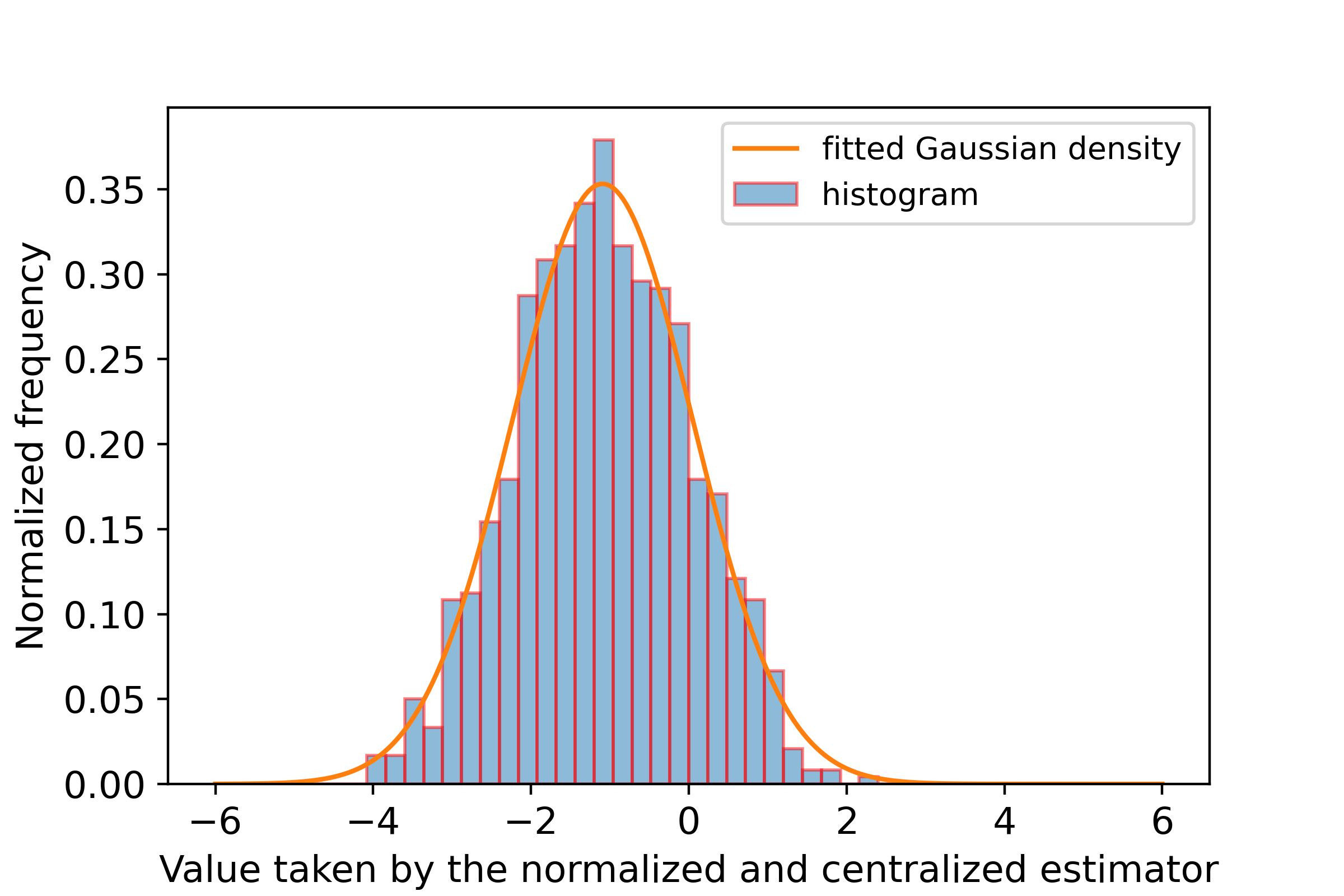}}
 \end{figure}

{\noindent\bf Additional Numerical Results of t-distribution}
The results of t-distribution we mentioned in numerical experiments are given by Figure \ref{fig:t_algo} and \ref{fig:t_naive}.
 
\textbf{Comparison with data splitting method}

In this part, we compare our method (noted as method 1) with a data splitting method (noted as method 2), where the method 2 uses $T^{1-\eta}$ number of samples for optimization only and the rest of samples for inference.

We provide some numerical results to illustrate the comparative performance of method 1 and method 2. We use the same setting in our numerical experiments and test for different values of $T$, and for method 2, we also test different choices of $\eta$. Recall that this parameter refers to using a small part $T^{1-\eta}$ number of samples for optimization and the rest for inference. We conduct experiments to implement the two methods and document two metrics $\mathbb{E}\Vert \hat{\theta}-\theta^*\Vert_2$ and $\mathbb{E}\vert \hat{f}(\hat{\theta})-f(\theta^*)\vert$. Here the first metric $\mathbb{E}\Vert \hat{\theta}-\theta^*\Vert_2$ is to evaluate the performance of the optimization task for a given method, as it measures how close the final $\hat{\theta}$ is to $\theta^*$. The second metric $\mathbb{E}\vert \hat{f}(\hat{\theta})-f(\theta^*)\vert$ is to evaluate the performance of the inference task on the estimated optimal objective function value, as it measures the how close the estimation $\hat{f}(\hat{\theta})$ is to $f(\theta^*)$.  

The results are drawn in following six tables (Table 1-6). All the results used 100 replications. Table 1 and Table 2 document the two metrics for comparing method 1 and method 2. The experiment setting uses Bernoulli distribution as the outcome distribution. To further examine the comparative performances, we uses different distribution as the outcome distribution, including Gaussian and Pareto. Specifically, Table 3 and Table 4 document the two metrics for comparing method 1 and method 2, with experiment setting using Gaussian distribution.  Table 5 and Table 6 document the two metrics for comparing method 1 and method 2, with experiment setting using Pareto distribution.

{In the tables, the value presented is the ratio of the error (method 2's error divided method 1's error). In Table 1, the first column represents the different settings of $T$ in the experiment. For example, 3E05 means $3\times 10^5$. The first row represents the different choices of $\eta$ by method 2 ($T^{1-\eta}$ number of samples for optimization and the rest for inference). For each entry, say the entry corresponding to $T = 3E05$ and $\eta = 0.1$ in Table 1, the value 1.40 in that entry represents the ratio 
}
\begin{align*}
    \frac{\text{Error } \mathbb{E} \Vert \hat{\theta}-\theta^*\Vert_2  \text{ incurred by method 2 by setting } \, \eta = 0.1 } {\text{Error } \mathbb{E} \Vert \hat{\theta}-\theta^*\Vert_2  \text{ incurred by method 1 }}
\end{align*}
in an experiment setting with $T = 3\times 10^5$.
{For Table 2, as an example, the entry corresponding to $T = 3E05$ and $\eta = 0.1$ shows value 1.26. Such value represents the ratio }
\begin{align*}
    \frac{\text{Error } \mathbb{E}\vert\hat{f}(\hat{\theta})-f(\theta^*)\vert  \text{ incurred by method 2 by setting } \, \eta = 0.1 } {\text{Error } \mathbb{E}\vert\hat{f}(\hat{\theta})-f(\theta^*)\vert  \text{ incurred by method 1 }}
\end{align*}
in an experiment setting with $T = 3\times 10^5$.

\begin{table}[h]
\caption{This table presents the ratio of optimization error $\mathbb{E} \Vert \hat{\theta}-\theta^*\Vert_2$ incurred by method 2 divided by error incurred by method 1. The parameter $\eta$ corresponds to the fraction of samples $T^{1-\eta}$ used for optimization by the method 2. This example uses the outcome distribution as Bernoulli.}
\begin{center}
\begin{tabular}{c|c|c|c|c|c|c}
\hline
$T$ &$\eta=0.025$   & $\eta=0.05$ &$\eta=0.1$ &$\eta=0.2$ &$\eta=0.4$  &$\eta=0.8$   \\ \hline
1E05  & 1.06 & 1.12 & 1.41 & 2.28 & 5.13 & 26.6
     \\ \hline

3E05  & 0.98 & 1.12 & 1.40 & 1.82 & 5.23 & 34.1 \\ \hline

1E06  & 0.96 & 1.01 & 1.25 & 2.21 & 5.75&45.3\\ \hline

\end{tabular}
\end{center}

\end{table}

\begin{table}[h!]
\caption{This table presents the ratio of inference error on function value $\mathbb{E}\vert\hat{f}(\hat{\theta})-f(\theta^*)\vert$ incurred by method 2 divided by error incurred by method 1. The parameter $\eta$ corresponds to the fraction of samples $T^{1-\eta}$ used for optimization by the method 2. This example uses the outcome distribution as Bernoulli.}
\begin{center}
\begin{tabular}{c|c|c|c|c|c|c}
\hline
$T$&$\eta=0.025$   & $\eta=0.05$ &$\eta=0.1$ &$\eta=0.2$ &$\eta=0.4$  &$\eta=0.8$   \\ \hline
1E05 & 1.97 & 1.67 & 1.36 & 1.35 & 3.66 & 74.1  \\ \hline

3E05& 1.99 & 1.54 & 1.26 & 1.20 & 3.82 & 104\\ \hline

1E06& 1.92 & 1.61 & 1.30 & 1.39 & 4.67&201\\ \hline

\end{tabular}
\end{center}

\end{table}

\begin{table}[h!]
\caption{This table presents the ratio of optimization error $\mathbb{E} \Vert \hat{\theta}-\theta^*\Vert_2$ incurred by method 2 divided by error incurred by method 1. The parameter $\eta$ corresponds to the fraction of samples $T^{1-\eta}$ used for optimization by the method 2. This example uses the outcome distribution as Gaussian ($N(f(\theta^*, 0.075^2))$).}
\begin{center}
\begin{tabular}{c|c|c|c|c|c|c}
\hline
$T$ &$\eta=0.025$   & $\eta=0.05$ &$\eta=0.1$ &$\eta=0.2$ &$\eta=0.4$  &$\eta=0.8$   \\ \hline
1E05  & 1.05 & 1.20 & 1.41 & 2.25 & 6.28 & 32.4
     \\ \hline

3E05  & 0.94 & 1.06 & 1.33 & 2.18 & 6.56 & 42.6 \\ \hline

1E06  & 0.91 & 1.25 & 1.57 & 2.84 & 8.03&68.4\\ \hline

\end{tabular}
\end{center}

\end{table}

\begin{table}[h!]
\caption{This table presents the ratio of inference error on function value $\mathbb{E}\vert\hat{f}(\hat{\theta})-f(\theta^*)\vert$ incurred by method 2 divided by error incurred by method 1. The parameter $\eta$ corresponds to the fraction of samples $T^{1-\eta}$ used for optimization by the method 2. This example uses the outcome distribution as Gaussian ($N(f(\theta^*, 0.075^2))$).}
\begin{center}
\begin{tabular}{c|c|c|c|c|c|c}
\hline
$T$&$\eta=0.025$   & $\eta=0.05$ &$\eta=0.1$ &$\eta=0.2$ &$\eta=0.4$  &$\eta=0.8$   \\ \hline
1E05 & 1.75 & 1.72 & 1.22 & 1.36 & 4.80 & 114  \\ \hline

3E05& 1.81 & 1.43 & 0.99 & 1.09 & 4.67 & 171\\ \hline

1E06& 1.78 & 1.41 & 1.26 & 1.37 & 5.32&294\\ \hline

\end{tabular}
\end{center}

\end{table}

\begin{table}[h!]
\caption{This table presents the ratio of optimization error $\mathbb{E} \Vert \hat{\theta}-\theta^*\Vert_2$ incurred by method 2 divided by error incurred by method 1. The parameter $\eta$ corresponds to the fraction of samples $T^{1-\eta}$ used for optimization by the method 2. This example uses the outcome distribution as Pareto.}
\begin{center}
\begin{tabular}{c|c|c|c|c|c|c}
\hline
$T$ &$\eta=0.025$   & $\eta=0.05$ &$\eta=0.1$ &$\eta=0.2$ &$\eta=0.4$  &$\eta=0.8$   \\ \hline
1E05  & 1.00 & 1.01 & 1.23 & 2.10 & 3.44 & 19.1
     \\ \hline

3E05  & 1.11 & 1.31 & 1.51 & 2.31 & 4.91 & 31.8 \\ \hline

1E06  & 0.85 & 1.03 & 1.25 & 2.06 & 4.95&33.3\\ \hline

\end{tabular}
\end{center}

\end{table}

\begin{table}[h!]
\caption{This table presents the ratio of inference error on function value $\mathbb{E}\vert\hat{f}(\hat{\theta})-f(\theta^*)\vert$ incurred by method 2 divided by error incurred by method 1. The parameter $\eta$ corresponds to the fraction of samples $T^{1-\eta}$ used for optimization by the method 2. This example uses the outcome distribution as Pareto.}
\begin{center}
\begin{tabular}{c|c|c|c|c|c|c}
\hline
$T$&$\eta=0.025$   & $\eta=0.05$ &$\eta=0.1$ &$\eta=0.2$ &$\eta=0.4$  &$\eta=0.8$   \\ \hline
1E05 & 1.74 & 1.33 & 1.01 & 1.19 & 1.98 & 39.8  \\ \hline

3E05& 1.94 & 1.54 & 1.22 & 1.33 & 2.79 & 88.5\\ \hline

1E06& 1.97 & 1.29 & 1.09 & 1.30 & 4.09&136\\ \hline

\end{tabular}
\end{center}
\end{table}

From above tables we can see that, when $\eta$ in method 2 is small, then the optimization task can have comparable or even slightly better performance with method 1, but method 2 is significantly worse than method 1 on the inference task. When $\eta$ becomes larger (meaning that fewer samples are used in the optimization step and more samples are used in the inference step), method 2 can become better on the inference task (though still not as good as method 1), but method 2's performance on the optimization task becomes comparatively worse. The results hold similar patterns for different choices of $T$. Overall, for the two tasks (optimization task and inference task), there are no scenarios where method 2 are better than method 1 in both the two metrics. Meanwhile, method 1 is usually better or not worse than method 2 in most scenarios. 

\end{document}